\documentclass[final,3p,times,twocolumn]{elsarticle}

\usepackage{amssymb}
\usepackage{amsmath}
\usepackage{dblfloatfix}
\usepackage{xcolor}
\usepackage{listings}
\usepackage{placeins}
\usepackage{hyperref}

\newcounter{bla}

\journal{Computer Physics Communications}
\makeatletter
\def\input@path{{./images/}}
\makeatother

\graphicspath{{./images/}}

\begin{document}

\begin{frontmatter}



\title{Eilmer: an open-source multi-physics hypersonic flow solver}


\author[a]{Nicholas N. Gibbons}
\author[a]{Kyle A. Damm}
\author[a]{Peter A. Jacobs}
\author[a]{Rowan J. Gollan\corref{author}}

\cortext[author] {Corresponding author.\\\textit{E-mail address:} r.gollan@uq.edu.au}
\address[a]{Centre for Hypersonics, School of Mechanical \& Mining Engineering, The University of Queensland}

\begin{abstract}
This paper introduces Eilmer, a general-purpose open-source compressible flow solver developed at the University of Queensland, designed to support research calculations in hypersonics and high-speed aerothermodynamics. Eilmer has a broad userbase in several university research groups and a wide range of capabilities, which are documented on the project's website, in the accompanying reference manuals, and in an extensive catalogue of example simulations. The first part of this paper describes the formulation of the code: the equations, physical models, and numerical methods that are used in a basic fluid dynamics simulation, as well as a handful of optional multi-physics models that are commonly added on to do calculations of hypersonic flow. The second section describes the processes used to develop and maintain the code, documenting our adherence to good programming practice and endorsing certain techniques that seem to be particularly helpful for scientific codes. The final section describes a half-dozen example simulations that span the range of Eilmer's capabilities, each consisting of some sample results and a short explanation of the problem being solved, which together will hopefully assist new users in beginning to use Eilmer in their own research projects.
\end{abstract}


\end{frontmatter}



\newpage

{\bf PROGRAM SUMMARY}

\begin{small}
\noindent
{\em Program Title:} Eilmer                                          \\
{\em CPC Library link to program files:} (to be added by Technical Editor) \\
{\em Developer's repository link:} https://github.com/gdtk-uq/gdtk \\
{\em Code Ocean capsule:} (to be added by Technical Editor)\\
{\em Licensing provisions(please choose one):} GPLv3 \\
{\em Programming language:} D, Lua                            \\
{\em Supplementary material:} https://gdtk.uqcloud.net  \\
{\em Nature of problem:}\\
Eilmer solves the compressible Navier-Stokes equations with a particular emphasis on
flows at hypersonic speeds.
The code includes modelling for high-temperature gas effects such as
chemical and vibrational nonequilibrium.
Eilmer can be used for the simulation for unsteady and steady flows.
\\
{\em Solution method:}\\
The code is implemented in D~\cite{1} and built on a finite-volume formulation
that is capable of solving the Navier-Stokes equations in 2D and 3D
computational domains, discretised with structured or unstructured grids.
Grids may be generated using a built-in parametric scripting tool or imported from commercial gridding software. 
The inviscid fluxes are computed using the reconstruction-evolution approach.
In structured-grid mode, reconstruction stencils up to fourth-order spatial
accuracy are available.
In unstructured-grid mode, least-squares reconstruction provides second-order
spatial accuracy.
A variety of flux calculators are available in the code.
Viscous fluxes are computed with compact stencils with second-order spatial accuracy.
For unsteady flows, explicit time-stepping with low-order RK-family schemes are available, along with a point-implicit Backward-Euler update scheme for stiff systems of equations.
For steady flows convergence can be greatly accelerated Jacobian-free Newton-Krylov update scheme, which seeks a global minimum in the residuals using a series of large pseudo-timesteps.
Domain decomposition is used for parallel execution using both shared memory and distributed memory
programming techniques.
\\
{\em Additional comments:}\\
Eilmer provides a programmable interface for pre-processing, post-processing
and user run-time customisations.
The programmable interface is enabled using a built-in embedded interpreter for the Lua programming language~\cite{2}.
Run-time customisations include used-defined boundary conditions, source terms
and grid motion.

\end{small}

\newpage

\section{Introduction}
\label{sec:intro}

Hypersonic flight is a frontier of aerodynamic science that presents us with many challenges. Some arise from the physical complexity of the fluid mechanics involved: the presence of shockwaves, turbulence, thermochemical nonequilibrium, radiation, and other phenomena that prevents us from developing a simple theory for the flow around a hypersonic vehicle. Pushing from the other direction are problems of operational requirements: the engineering and practical considerations that must be solved to build an efficient flying machine that can survive the extreme conditions, and with enough mass to spare for a useful payload.

In spite of these challenges, recent successes in hypersonic flight have shown that the challenges can be overcome, thanks in part to the testing, simulation, and design tools accumulated by the research community over many decades of work. These tools include flight testing of expendable models, ground testing in hypersonic wind tunnels, and computer modelling and simulation, chiefly, the use of computational fluid dynamics (CFD).

The role of CFD in the study of hypersonic flows is now ubiquitous.
CFD is used in support of ground testing, flight testing, and, increasingly, on its own as a numerical wind tunnel.
In ground testing, CFD is used in the design and analysis of test articles,
and in the design and analysis of the test facilities themselves.
For example, hypersonic test facilities typically cannot directly measure everything about the test flow they produce, and rely on CFD to reconstruct the full flow conditions delivered during the experiment.
Flight testing is performed infrequently due to the time, expense and risk involved.
When flight testing is undertaken, CFD is used in the design stage and as part of
proving flight-readiness.
With advances in the accuracy and robustness of CFD algorithms and the power of computers,
it is now possible to resolve all time and length scales of a wall-bounded turbulent hypersonic flow.
When used in this manner, CFD becomes a numerical wind tunnel, and, as such, a tool for investigation of flow physics.
This mode of CFD usage is quite different from engineering analysis, which aims to support decision making.
This list is just a small sample of the roles for CFD in advancing hypersonic vehicle design.

The varied roles for CFD in the study of hypersonic flow present a challenge for the development of a CFD tool.
The strategies for addressing this challenge fall broadly into two approaches: either
develop a niche code to tackle one application; or develop a comprehensive code that has a broad range of capabilities but less specificity.
The advantages of the niche code approach include: quick development time; limited feature set to 
verify and validate; use of the best algorithms and computer architectures for the application;
and a relatively small surface area for code maintenance.
The principal disadvantage of this approach, the niche bespoke code, is one of reusing the code
for new applications.
As such, the niche code approach can face limitations in versatility such as restricted sets of boundary conditions,
limitations on geometric complexity, and limitations on physical modelling capability.
In some cases, these codes cannot be easily modified for new applications because the code is so highly
customised towards its original purpose.

At the other extreme, the strategy involves building a comprehensive code that serves multiple purposes.
The advantages of comprehensive code development include: that developers and users learn one code base;
synergies in code re-use for varied applications (such as I/O, message passing, geometry routines and gas modelling);
and relative ease to add new modelling capability not originally envisaged.
The disadvantages when contrasted to the niche code are:
long development time; large modelling capability that requires verification and validation
(see Kleb and Wood~\cite{kleb_2004_cfd} for a discussion of this issue);
trade-offs on algorithm choices against the need for generalisation; and
a large surface area of code for maintenance.

These approaches to development fall on a continuum, and, likely, any comprehensive
code began life with a niche purpose.
We are not advocating one approach over the other here as each has their merits
relative to circumstances.
For example, in academia, the niche code approach makes a lot of sense in a
world of short funding cycles.
The niche code is also extremely useful when developing a completely novel algorithm.
By contrast, the comprehensive code approach is better suited to government and
industry institutions where funding time scales are typically longer.
We present this view on simulation code development approaches to give context
to the work presented in this paper.
Here we describe the simulation code Eilmer.
It is a comprehensive code for hypersonic multi-physics simulation that
has been developed for 30 years and, principally, that development
has been at The University of Queensland.

The advantages and disadvantages as described are mostly concerns for the developer.
When viewed from a user's perspective, the benefits of a comprehensive code are clearer.
These benefits are a reduction in the investment learning-curve time by using one tool for a variety of applications;
and a larger community of users for support.
As mentioned earlier, a hypersonics engineer has to use CFD in a variety of situations at varied levels
of modelling fidelity.
So, there is good motivation for a comprehensive simulation code for hypersonic flows.

The purpose of this paper is to introduce Eilmer as a comprehensive hypersonic
flow solver with multi-physics capability.
The timing of this paper coincides with a version 4.0 release of the code.
We wish to show that there are few open-source CFD tools with hypersonic
multi-physics simulation capability.
We also aim to demonstrate that Eilmer is supported for general use
with extensive documentation.
Furthermore, modelling capability can be augmented by users in two ways:
by user-defined runtime customisations written in the scripting language Lua;
or by direct modification to the source code.

%
%

There has been other recent activity in the development of open-source
flow solvers for the compressible flow regime.
We restrict our attention to open-source offerings since we believe
these offer the best opportunity for code re-use and re-purpose.
The STREAmS code~\cite{bernardini_2021_streams} is a specialised
solver for performing direct numerical simulation on wall-bounded
compressible flows.
The code is specialised both in terms of algorithms and computer
architectures. It is designed for high-resolution calculations
and achieves this by demonstrating excellent scaling on multi-GPU
HPC platforms (up to 1024 GPUs).
STREAmS is restricted to Cartesian domains.
For treating geometric complexity,
Romero et al.\ have released ZEFR~\cite{romero_2020_zefr} as an
open-source code to simulate compressible viscous flows.
ZEFR is also targeted towards high-resolution capability and
the associated computational demands are met by developing
for multi-GPU architectures, as for STREAmS.
OpenSBLI~\cite{lusher_2021_opensbli} is another open-source code that
falls into the category of scale-resolving flow codes, accelerated
by GPUs.
OpenSBLI is actually a code-generation system to produce code
for heterogeneous architectures on modern HPC platforms.
In terms of algorithms, it is based on high-order finite
differences on structured curvilinear grids, with both WENO
and TENO schemes for shock-capturing.
All of these codes simulate ideal gases.
This is a limitation for application in the high-temperature
hypersonic flow regimes.
By contrast, the HTR solver~\cite{direnzo_2020_htr} by Di Renzo et al.\
include modelling of thermochemical nonequilibrium effects typical in hypersonic flows.
HTR is also aimed at very high-resolution calculations for DNS work,
and is similarly accelerated by implementation for multi-GPU architectures.
All of these codes could be considered quite niche in terms of application.

The hy2Foam code~\cite{casseau_2016_two} is an example of an
open-source hypersonic flow solver that is quite general purpose.
Built on top of OpenFOAM, hy2Foam has access to numerical discretization
algorithms to treat a variety of geometric complexity.
The hy2Foam authors added physical modelling in terms of nonequilibrium
thermochemical effects so that the code can be applied in the
high-temperature flow regime.
Another open-source project for multi-physics simulation in the
hypersonic flow regime is SU2-NEMO~\cite{maier_2021_su2nemo}.
This code, built as an extension to SU2~\cite{economon_2016_su2},
is an unstructured solver so that objects with high geometric complexity can be simulated.
It has modelling for high-temperature nonequilibrium effects,
such as finite-rate chemistry and two-temperature thermal relaxation,
which are required for providing engineering estimates of
flow field quantities in high Mach number flight.

The Eilmer code, presented in this paper, is a comprehensive
open-source code for multi-physics simulation of hypersonic flows
fit for multiple purposes.
We believe this range of physical modelling capability, combined with the open-source nature and active development of the project, makes a novel contribution to the academic
CFD simulation community.
Eilmer offers capabilities for time accurate simulation and accelerated steady-state
analyses.
Computational domains of considerable geometric complexity are handled with
multiple-block structured or unstructured grids.
For simulating high-temperature effects, the gas models include
mixtures of thermally perfect gases, multi-temperature Boltzmann gases,
and state-to-state models for extreme nonequilibrium.
The code includes modelling of the nonequilibrium relaxation
via finite-rate chemistry modules and thermal energy exchange models.
Multi-physics simulation capability includes fluid-structure interactions,
fluid-thermal interactions, and fluid-radiation coupling.
Aerodynamic shape optimisation with many design parameters
is enabled by an adjoint-solver.
Most of these capabilities can be customised for special purpose
via user-defined runtime controls.

In this paper, we present the Eilmer code showing what
it offers to the community in terms of application and re-use.
In Section~\ref{sec:formulation}, the core numerical formulation
based on a finite-volume approach is presented.
To build trust in the quality and longevity of the code,
we discuss the development process in Section~\ref{sec:development},
including quality control via verification and validation.
Section~\ref{sec:scaling} presents the parallelisation algorithm used to
run simulations on distributed computational hardware, and includes some
test data to prove good scaling performance up to 1000 physical cores.
We have mentioned that Eilmer supports a variety of use cases.
In Section~\ref{sec:applications},
we present examples of those uses cases: engineering analysis,
optimised aerodynamic design, and flow physics investigation.
An unusual aspect of Eilmer as a research code is that it
is used in the classroom;
we include an example of student simulations in
Section~\ref{sec:applications} also.

\section{Formulation}
\label{sec:formulation}
The core of Eilmer is a set of numerical routines for solving mixed hyperbolic/parabolic partial differential equations (PDEs) that have been discretised on a grid of finite-volume cells. These PDEs are the compressible Navier-Stokes equations for viscous flow in two or three dimensions, augmented with various add-ons, simplifications, and optional extras for including additional physics. Rather than exhaustively document the solver's entire suite of capabilities, in this section we present the core flow equations and the foundational assumptions underlying them, followed by three of the most common multi-physics capabilities that are commonly used in hypersonics research calculations. Additional details about any of these topics can be found in the code's online \href{https://gdtk.uqcloud.net/docs/all-the-docs/about}{documentation} or the cited references.\\

\subsection{The Navier-Stokes Equations}
\label{fluid_equations}
Equation \ref{integral_form} summarises the fundamental set of conservation equations of compressible fluid flow, expressed in conservative, integral form. They consist of a vector of conserved quantities $\mathbf{U}$ that is tracked inside of a small volume $V$, which changes in time as each quantity flows through the surfaces of the volume, as determined by the fluxes $\mathbf{F}$, or generated internally by a source term $\mathbf{Q}$.

\begin{equation}
\frac{\partial}{\partial t} \int_V \mathbf{U}~dV = - \oint_S (\mathbf{F}_i - \mathbf{F}_v) \cdot \hat{n}~ dA + \int_V \mathbf{Q}~dV
\label{integral_form}
\end{equation}

Embodied in these equations are two foundational assumptions that underlie all simulations performed with Eilmer. The first is that the flow is modelled as a \emph{continuum}: A field with values at each point in space rather than a discrete collection of molecules. The second is that the fields are \emph{compressible}, in the sense that information about a disturbance must travel in waves at a finite speed from the point of the disturbance. The first assumption is valid when the particles making up the fluid are so small and so numerous that they form a well behaved statistical ensemble, though the assumption can break down in flows that are very cold, very low density, or interacting with a very small object. The second assumption is appropriate to flows of any velocity, although a low speed flow is much more efficiently modelled by an incompressible formulation where the signal speed can be approximated as infinitely fast. 

For a nonreacting flow in three-dimensions, with no additional physics present, the conserved quantities are the mass, x-momentum, y-momentum, z-momentum, and energy --- or more precisely, the density per unit volume of these quantities.

\begin{equation}
\mathbf{U} = 
\begin{bmatrix}
    \rho    \\
    \rho u  \\
    \rho v  \\
    \rho w  \\
    \rho E  \\
\end{bmatrix}
\end{equation}

The inviscid flux vector $\mathbf{F}_i$ tracks the movement of conserved quantities due to bulk fluid motion, and consists of elements that are each in turn vectors in three-dimensional space. 

\begin{equation}
\mathbf{F}_i = 
\begin{bmatrix}
    \rho u  \\
    \rho u^2 + p  \\
    \rho v u \\
    \rho w u \\
    \rho E u + p u \\
\end{bmatrix}
\hat{i}
+
\begin{bmatrix}
    \rho v  \\
    \rho u v   \\
    \rho v^2 + p \\
    \rho w v \\
    \rho E v + p v \\
\end{bmatrix}
\hat{j}
+
\begin{bmatrix}
    \rho w  \\
    \rho u w \\
    \rho v w \\
    \rho w^2 + p \\
    \rho E w + p w \\
\end{bmatrix}
\hat{k}
\end{equation}

These fluxes are distinguished from the viscous fluxes, which arise from nano-scale disordered motion of the particles making up the fluid, and tend to involve derivatives of the primitive quantities:

\begin{equation}
\begin{split}
\mathbf{F}_v = 
\begin{bmatrix}
    0  \\
    \tau_{xx}  \\
    \tau_{yx} \\
    \tau_{zx} \\
    \tau_{xx} u + \tau_{yx} v +\tau_{zx} w + q_x \\
\end{bmatrix}
\hat{i}
+
\begin{bmatrix}
    0  \\
    \tau_{xy}  \\
    \tau_{yy} \\
    \tau_{zy} \\
    \tau_{xy} u + \tau_{yy} v +\tau_{zy} w + q_y \\
\end{bmatrix}
\hat{j} \\
+
\begin{bmatrix}
    0  \\
    \tau_{xz}  \\
    \tau_{yz} \\
    \tau_{zz} \\
    \tau_{xz} u + \tau_{yz} v +\tau_{zz} w + q_z \\
\end{bmatrix}
\hat{k}
\end{split}
\end{equation}

\noindent where the viscous stress tensor $\tau$ is composed from derivatives of the velocity vector, the viscosity $\mu$ and the volume (or bulk) viscosity $\lambda$.

\begin{equation}
\begin{split}
\tau_{xx} = 2 \mu \frac{\partial u}{\partial x} + \lambda \left(\frac{\partial u}{\partial x} + \frac{\partial v}{\partial y} + \frac{\partial w}{\partial z} \right)\\
\tau_{yy} = 2 \mu \frac{\partial v}{\partial y} + \lambda \left(\frac{\partial u}{\partial x} + \frac{\partial v}{\partial y} + \frac{\partial w}{\partial z} \right)\\
\tau_{zz} = 2 \mu \frac{\partial w}{\partial z} + \lambda \left(\frac{\partial u}{\partial x} + \frac{\partial v}{\partial y} + \frac{\partial w}{\partial z} \right)\\
\tau_{xy} = \tau_{yx} = \mu \frac{\partial u}{\partial y} + \frac{\partial v}{\partial x}\\
\tau_{yz} = \tau_{zy} = \mu \frac{\partial v}{\partial z} + \frac{\partial w}{\partial y}\\
\tau_{xz} = \tau_{zx} = \mu \frac{\partial u}{\partial z} + \frac{\partial w}{\partial x}
\end{split}
\end{equation}

The code generally assumes that $\lambda$ can be evaluated using Stokes' hypothesis as $\lambda = -\frac{2}{3}\mu$. The heat conduction vector $q$ is modelled using Fourier's law and the gradient of the temperature $T$, with conductivity $\kappa$.
\begin{equation}
\begin{split}
q_{x} = \kappa \frac{\partial T}{\partial x}\\
q_{y} = \kappa \frac{\partial T}{\partial y}\\
q_{z} = \kappa \frac{\partial T}{\partial z}
\end{split}
\end{equation}

The transport properties $\mu$ and $\kappa$ can be computed by any of several of different models depending the needs of the user. For compressible flows with simple thermochemistry Sutherland's approximation can be used, but for higher temperatures or complex mixtures of species the user can request the curve-fits from the NASA Glenn thermodynamic database \cite{glennthermo}, which includes a comprehensive database of individual species. For the thermal equation of state we generally assume perfect gas behaviour, meaning that the internal energy is function of temperature only. Under this assumption, the pressure $p$ can be computed using the temperature, total density, and the specific gas constant $R$.

\begin{equation}
p = \rho R T
\end{equation}

The last piece of the formulation is an expression for the total energy $E$, which consists of a kinetic energy component from the gas velocity and an internal energy $e$ that is a function of temperature.

\begin{equation}
\rho E = \frac{1}{2} \rho \left(u^2 + v^2 + w^2 \right) + \rho ~ e(T)
\end{equation}

The relationship between internal energy $e$ and temperature can also be specified in different ways, including an ideal gas assumption that is fast but valid only for flow below $<500 K$, and a thermally perfect model with Glenn database curve-fits \cite{glennthermo} that are usually good up to $\approx 20,000 K$.

\subsection{Optional Extras for Hypersonic Multi-physics}
\label{multi_physics}
Hypersonic flows exhibit a wide range of complicated physical phenomena. High energies in the gas state can cause chemical reactions or nonequilibrium thermodynamics, both of which commonly progress on similar timescales to the flow itself and lead to time-dependant thermal behaviour that is difficult to model analytically. Another complication is turbulence: Three-dimensional, unsteady, fine scale fluctuations in the fluid properties that occur unavoidably at a high enough Reynolds number. These effects require modifying the fluid-mechanic formulation above into a coupled or ``multi-physics'' approach, with extra equations added to the set of hyperbolic PDEs. In this section we present the three most common extra physics models, starting with those needed for simulating flows with chemical reactions.\\

Both exothermic reactions like combustion and endothermic reactions like air dissociation are commonly simulated with Eilmer, which requires the introduction of an extra transport equation for the partial density of each chemical species $\rho_s$, where the index $s$ runs across every species in a given reaction mechanism.

\begin{equation}
\mathbf{U} = 
\begin{bmatrix}
    \vdots  \\
    \rho_1  \\
    \rho_2  \\
    \vdots  \\
    \rho_s  \\
\end{bmatrix}
\end{equation}

Each species is convected with the flow, diffused via concentration gradients, and created or destroyed by chemical reactions. These effects appear in the inviscid, viscous, and source vectors respectively. 

\begin{equation}
\mathbf{F}_i = 
\begin{bmatrix}
    \vdots  \\
    \rho_1 u \\
    \rho_2 u \\
    \vdots  \\
    \rho_s u \\
\end{bmatrix}
\hat{i}
+
\begin{bmatrix}
    \vdots  \\
    \rho_1 v \\
    \rho_2 v \\
    \vdots  \\
    \rho_s v \\
\end{bmatrix}
\hat{j}
+
\begin{bmatrix}
    \vdots  \\
    \rho_1 w \\
    \rho_2 w \\
    \vdots  \\
    \rho_s w \\
\end{bmatrix}
\hat{k}
\end{equation}

\begin{equation}
\mathbf{F}_v = 
\begin{bmatrix}
    \vdots  \\
    \rho D_1 \frac{\partial Y_1}{\partial x} \\
    \rho D_2 \frac{\partial Y_2}{\partial x} \\
    \vdots  \\
    \rho D_s \frac{\partial Y_s}{\partial x} \\
\end{bmatrix}
\hat{i}
+
\begin{bmatrix}
    \vdots  \\
    \rho D_1 \frac{\partial Y_1}{\partial y} \\
    \rho D_2 \frac{\partial Y_2}{\partial y} \\
    \vdots  \\
    \rho D_s \frac{\partial Y_s}{\partial y} \\
\end{bmatrix}
\hat{j}
+
\begin{bmatrix}
    \vdots  \\
    \rho D_1 \frac{\partial Y_1}{\partial z} \\
    \rho D_2 \frac{\partial Y_2}{\partial z} \\
    \vdots  \\
    \rho D_s \frac{\partial Y_s}{\partial z} \\
\end{bmatrix}
\hat{k}
\end{equation}

In these expressions, $Y_s$ and $D_s$ are the mass fraction and mixture-averaged diffusion coefficient of species $s$. The diffusion of the species also introduces a new term into the energy transport equation, to account for the latent heat of formation tied up in the specific enthalpy of each species $h_s$. This alters the energy equation's entry in the viscous flux vector:

\begin{equation}
\begin{aligned}
  & F_v^{\rho E} =\\
  & (\tau_{xx} u + \tau_{yx} v +\tau_{zx} w + q_x - \sum_s \rho h_s D_s \frac{\partial Y_s}{\partial x}) ~ \hat{i} \\
+ & (\tau_{xy} u + \tau_{yy} v +\tau_{zy} w + q_y - \sum_s \rho h_s D_s \frac{\partial Y_s}{\partial y}) ~ \hat{j} \\
+ & (\tau_{xz} u + \tau_{yz} v +\tau_{zz} w + q_z - \sum_s \rho h_s D_s \frac{\partial Y_s}{\partial z}) ~ \hat{k}
\end{aligned}
\end{equation}

The source term $\mathbf{Q}$ is usually zero for the fluid equations alone, although Eilmer does contain an axisymmetric mode that includes geometric source terms that account for the complications of a polar co-ordinate system. But the source terms are also used heavily in almost all of the multiphysics models. For chemical reactions they consist of the net mass density production rates for each species.
\begin{equation}
\mathbf{Q} = 
\begin{bmatrix}
    \vdots  \\
    M_1 \dot{\omega}_1\\
    M_2 \dot{\omega}_2\\
    \vdots  \\
    M_s \dot{\omega}_s\\
\end{bmatrix}
\end{equation}

These rates are assembled by summing over the reactions in a reaction mechanism $r$.
\begin{equation}
\dot{\omega}_s =  \sum_r -\nu_{rs} \left[k_{r}^f \Pi_n [X_n]^{\nu_{rn}^f} - k_{r}^b \Pi_n [X_n]^{\nu_{rn}^b}\right]
\label{omega_s}
\end{equation}

\noindent where $\nu^f_{rs}$ and $\nu^b_{rs}$ are the forward and backward stoichiometric coefficients for the species $s$ in reaction $r$, $\nu_{rs}$ is the net stoichiometric coefficient $\nu_{rs} = \nu^f_{rs} - \nu^b_{rs}$, $[X_n]$ is the molar concentration of species $n$, and $k^f_r$ and $k^b_r$ are the forward and backward reaction tick rates for reaction $r$. (This nomenclature is similar to that presented by \citet{anderson_hypersonics}, which should be consulted for further detail).

Usually $k^f$ is specified by an Arrhenius equation with coefficients $A$, $n$, and $C$ curve-fitted from reaction rate data, each reaction in the set $r$ having different coefficients.

\begin{equation}
k^f = A T^n e^{-C/T}
\end{equation}

The backward rates $k^b$ can also be specified in the same way, or they can be computed from the equilibrium coefficient for the given reaction $K_{eq}$

\begin{equation}
k^b = \frac{k^f}{K_{eq}} ~~~;~~~ K_{eq} = \left(\frac{p^{\circ}}{R_u T}\right)^{\sum_s \nu_s} \times exp\left(\frac{\sum_s \nu_{s} g_s^{\circ}(T) M_s}{R_u T}\right)
\end{equation}

\noindent where $g^{\circ} = g - T s^{\circ}$ is the Gibbs free energy per unit mass at standard pressure $p^{\circ}$, and $M_s$ is the molecular mass of species $s$. The final change for chemically reacting flows involves the energy equation and equation of state, which are assembled out of the partial densities, the species internal energies $e_s$, and the standard-state heats of formation $h^f_s = h_s(T=298.15)$, with 298.15 Kelvin chosen as the standard temperature. 

\begin{equation}
p = \sum_s \rho_s R_s T
\end{equation}

\begin{equation}
\rho E = \frac{1}{2} \rho \left(u^2 + v^2 + w^2 \right) + \sum_s \rho_s e_s(T) + \sum_s \rho_s h^f_s
\end{equation}

Another common phenomenon present in hypersonic flows is turbulence. Turbulence is the result of cascading instabilities in the flow that break down large disturbances in the flow into smaller ones, producing an unsteady fractal of swirling motion with a wide range of scales that is difficult to resolve on a finite-volume grid. A straightforward and reasonably successful modelling strategy to avoid this problem is Reynolds-Averaging, where the turbulent structures are time-averaged out of the simulation and accounted for with modified transport properties such as the turbulent viscosity $\mu_t$ and turbulent thermal conductivity $k_t$. Eilmer is equipped for Reynolds-Averaged Navier-Stokes (RANS) style turbulence modelling using various models that have been well-validated by the aerospace community, any of which add a number of extra transport equations to the formulation. As an example, the flux vectors for the one-equation Spalart-Allmaras model \cite{allmaras12} are shown below, which adds a conserved quantity for the turbulent viscosity $\rho \hat{\nu}$, an inviscid flux, a viscous flux, and a complicated source term. The two equation $k-\omega$ model \cite{wilcoxturbulence} and the high-fidelity Improved Delayed Detached Eddy Simulation \cite{wmlesspalart} technique are also available. 

\begin{equation}
\mathbf{U} = 
\begin{bmatrix}
    \vdots  \\
    \rho \hat{\nu}  \\
\end{bmatrix}
\end{equation}

\begin{equation}
\mathbf{F}_i = 
\begin{bmatrix}
    \vdots  \\
    \rho \hat{\nu} u \\
\end{bmatrix}
\hat{i}
+
\begin{bmatrix}
    \vdots  \\
    \rho \hat{\nu} v \\
\end{bmatrix}
\hat{j}
+
\begin{bmatrix}
    \vdots  \\
    \rho \hat{\nu} w \\
\end{bmatrix}
\hat{k}
\end{equation}

\begin{equation}
\mathbf{F}_v = 
\begin{bmatrix}
    \vdots  \\
    \frac{\rho (\nu + \hat{\nu})}{\sigma} \frac{\partial \hat{\nu}}{\partial x} \\
\end{bmatrix}
\hat{i}
+
\begin{bmatrix}
    \vdots  \\
    \frac{\rho (\nu + \hat{\nu})}{\sigma} \frac{\partial \hat{\nu}}{\partial y} \\
\end{bmatrix}
\hat{j}
+
\begin{bmatrix}
    \vdots  \\
    \frac{\rho (\nu + \hat{\nu})}{\sigma} \frac{\partial \hat{\nu}}{\partial z} \\
\end{bmatrix}
\hat{k}
\end{equation}


\begin{equation}
\mathbf{Q} = 
\begin{bmatrix}
    \vdots  \\
\rho c_{b1} (1 - f_{t2}) \hat{S} \hat{\nu} -
\rho \left[
    c_{w1} f_w - \frac{c_{b1}}{\kappa^2} f_{t2}
 \right] \left(\frac{\hat{\nu}}{d}\right)^2 + \\
\frac{1}{\sigma} \frac{\partial}{\partial x_j} \left(
    \rho (\nu + \hat{\nu}) \frac{\partial \hat{\nu}}{\partial x_j}
\right)  + \frac{c_{b2}}{\sigma} \rho \frac{\partial \hat{\nu}}{\partial x_i}\frac{\partial     \hat{\nu}}{\partial x_i}
\end{bmatrix}
\end{equation}

These equations introduce the kinematic viscosity $\nu$, and allow for the solution of the Spalart-Allmaras field variable $\hat{\nu}$ throughout the flow, using a large number of additional constants and functions that are defined in \citet{allmaras12}. The actual turbulent viscosity $\mu_t$ is then computed as:
\begin{equation}
\mu_t = \overline{\rho} \hat{\nu} \frac{ (\hat{\nu}/\nu)^3 }{ (\hat{\nu}/\nu)^3 + c_{v1}^3}
\end{equation}
and the turbulent conductivity $\kappa_t$ and diffusivity $D_t$ from the turbulent Prandtl number $\text{Pr}_t$ and turbulent Schmidt number $\text{Sc}_t$.
\begin{equation}
\kappa_t = \frac{c_p \mu_t}{\text{Pr}_t} ~~~ D_t = \frac{\mu_t}{\rho \text{Sc}_t}
\end{equation}

These transport properties are then added to their laminar equivalents, and the sum is used in place of $\mu$, $\kappa$, and $D_s$ in the preceding expressions.\\

The last commonly encountered multi-physics model of interest is thermal nonequilibrium. This phenomenon occurs during rapid expansions or compressions of the flow, such as behind a shockwave, and affects the statistical distribution of the molecules making up the gas. At the molecular scale, the gas particles are rapidly churning through different internal states of translational velocity as they drift and collide, but the average properties of the entire collection stays more-or-less constant. This average state is well-described by a statistical function called the Boltzmann distribution, which is parameterised by the temperature. A sudden compression can instantaneously change the distribution of translational velocities, but the energy of the gas is also stored in other modes, such as rotation or vibration of the molecules, which may take a finite amount of time to catch up as energy is transferred from the translational mode via collisions. During this time the thermodynamic state of the gas is in a time-dependent state referred to as thermal-nonequilibrium, and the governing equations need to be modified to account for the different relationship between pressure, energy, and temperature.

The starting point for this modification is to include an extra transport equation for each additional energy mode beyond the first. These equations track the convection and diffusion of energy in each specific mode, and have source terms describing the energy exchange with the other modes via collisions. For the purposes of illustration, this section describes the commonly applied Two-Temperature model, in which the translational and rotational energies of all the species are combined into one thermal mode, and the vibration and electronic energies of all the species are combined into a second mode. The vibrational/electronic energy $e_v$ is then governed by the following equations. 

\begin{equation}
\mathbf{U} = 
\begin{bmatrix}
    \vdots  \\
    \rho e_v  \\
\end{bmatrix}
\end{equation}

\begin{equation}
\mathbf{F}_i = 
\begin{bmatrix}
    \vdots  \\
    \rho e_v u \\
\end{bmatrix}
\hat{i}
+
\begin{bmatrix}
    \vdots  \\
    \rho e_v v \\
\end{bmatrix}
\hat{j}
+
\begin{bmatrix}
    \vdots  \\
    \rho e_v w \\
\end{bmatrix}
\hat{k}
\end{equation}

\begin{equation}
\begin{split}
\mathbf{F}_v = 
\begin{bmatrix}
    \vdots  \\
    \kappa_v \frac{\partial T_v}{\partial x} - \sum_s \rho h_{vs} D_s \frac{\partial Y_s}{\partial x} \\
\end{bmatrix}
\hat{i}\\
+
\begin{bmatrix}
    \vdots  \\
    \kappa_v \frac{\partial T_v}{\partial y} - \sum_s \rho h_{vs} D_s \frac{\partial Y_s}{\partial y} \\
\end{bmatrix}
\hat{j}\\
+
\begin{bmatrix}
    \vdots  \\
    \kappa_v \frac{\partial T_v}{\partial z} - \sum_s \rho h_{vs} D_s \frac{\partial Y_s}{\partial z} \\
\end{bmatrix}
\hat{k}
\end{split}
\end{equation}

\begin{equation}
\mathbf{Q} = 
\begin{bmatrix}
    \vdots  \\
    \sum_s Q^{vib-trans} + \sum_s Q^{elec-trans} + \sum_r Q^{vib-chem}_{r}
\end{bmatrix}
\label{Q_ee}
\end{equation}

With the vibrational/electronic energies of each species lumped together, the source term in this case consists of a simple sum over all the energy exchange mechanisms. Exchanges between the translational energy and the vibrational modes are typically modelled using the Landau-Teller relaxation formula, with a relaxation time for each species $\tau_s$ computed using the semi-empirical correlations of \citet{relaxation_millikan63} or, optionally, by custom-fitted constants.
\begin{equation}
Q^{vib-trans}_s = \rho \frac{e_v - e_v^*(T)}{\tau_s}
\end{equation}

At higher velocities such as those experienced by reentering spacecraft, the flow may be ionised by the large amounts of energy present in the flow. The electrons freed by this process are sometimes assumed to have their own energy mode, or they may be lumped in with the vibrational/electronic mode as above. Energy exchange from collisions between the free-electrons translational energy and the heavy-particle translational energy can be modelled using an expression from \citet{gnoffo_1989}, which includes curve-fitted data for the electron/heavy collision frequencies $\nu_{es}$.
\begin{equation}
Q^{elec-trans}_s = 2 \rho_{e-} \frac{3}{2} R_u (T - T_v) \sum_s \frac{\nu_{es}}{M_s}
\end{equation}

\noindent where $\rho_{e-}$ is the partial density of free electrons. The final term in expression \ref{Q_ee} concerns energy exchanged between modes by chemical reactions. We have implemented a model based on the formulation in \citet{knab_chem_tvib} to account for this effect, which can be written as follows by summing over the molecular dissociation reactions $r$, 
\begin{equation}
\begin{aligned}
& Q^{vib-chem}_{r} = \\
& k_{r}^f \Pi_n [X_n]^{\nu_{rn}^f} \times \left[ \frac{R_u \Theta_s}{exp(\Theta_v/T^{*}) -1} - \frac{D}{exp(D/R_u/T^{*}) - 1} \right] \\
- & k_{r}^b \Pi_n [X_n]^{\nu_{rn}^b} \frac{1}{2} \left[D - \Theta_v R_u \right] &
\end{aligned}
\end{equation}

\noindent where $\Theta_v$ is the characteristic vibrational temperature of the relevant molecule, $D$ is the dissociation energy liberated by the reaction, and $k_r^f$ and $k_r^b$ are the forward and backward chemical reaction rates, as defined in equation \ref{omega_s}.

\subsection{Numerical Discretisation}
\label{numerical_discretisation}
Together the equations presented in sections \ref{fluid_equations} and \ref{multi_physics} form a coupled set of partial differential equations in three dimensions, which are not analytically solvable except for a few special cases. However, by replacing with the differential terms with their discrete equivalents, it is possible to produce powerful and nearly exact numerical solutions of the equations on a digital computer. The first step in this process is to recast the continuous integral form of the conservation equations into a form based on small but finite-sized polyhedra called cells.
\begin{equation}
\frac{\partial \mathbf{U}}{\partial t} = - \frac{1}{V} \sum_f^{faces} (\mathbf{F}_i - \mathbf{F}_v)_f \cdot \hat{n}_f ~ A_f + \mathbf{Q}
\label{semidiscrete}
\end{equation}

This form replaces the boundary integral with a sum over the faces of the cell $f$, and the fluxes with the equivalent averages over the face in question. Likewise the conserved quantities $\mathbf{U}$ and the source terms $\mathbf{Q}$ are the volume averages of the state inside the cell. The flow field thus consist of a collection of small cells, each containing numbers for the pressure, temperature, density, flow velocity, and any other quantity of interest within. These numbers are then advanced in time through a sequence of discrete steps by evaluating equation \ref{semidiscrete} over and over again, each step broken up into the following substeps.

\begin{enumerate}
\item Reconstruct the flow on both sides of each face
\item Compute the inviscid fluxes $\mathbf{F}_i$
\item Compute the gradients of the flow at each face
\item Compute the viscous fluxes $\mathbf{F}_v$
\item Compute the source term $\mathbf{Q}$ in each cell
\item Compute the residual $\mathbf{R} = \partial \mathbf{U}/ \partial t$
\item Advance the conserved quantities in time
\end{enumerate}

\begin{figure}[h]
    \centering
    \def\svgwidth{0.95\columnwidth}
\begingroup%
  \makeatletter%
  \providecommand\color[2][]{%
    \errmessage{(Inkscape) Color is used for the text in Inkscape, but the package 'color.sty' is not loaded}%
    \renewcommand\color[2][]{}%
  }%
  \providecommand\transparent[1]{%
    \errmessage{(Inkscape) Transparency is used (non-zero) for the text in Inkscape, but the package 'transparent.sty' is not loaded}%
    \renewcommand\transparent[1]{}%
  }%
  \providecommand\rotatebox[2]{#2}%
  \newcommand*\fsize{\dimexpr\f@size pt\relax}%
  \newcommand*\lineheight[1]{\fontsize{\fsize}{#1\fsize}\selectfont}%
  \ifx\svgwidth\undefined%
    \setlength{\unitlength}{198.44192693bp}%
    \ifx\svgscale\undefined%
      \relax%
    \else%
      \setlength{\unitlength}{\unitlength * \real{\svgscale}}%
    \fi%
  \else%
    \setlength{\unitlength}{\svgwidth}%
  \fi%
  \global\let\svgwidth\undefined%
  \global\let\svgscale\undefined%
  \makeatother%
  \begin{picture}(1,0.39574558)%
    \lineheight{1}%
    \setlength\tabcolsep{0pt}%
    \put(0,0){\includegraphics[width=\unitlength,page=1]{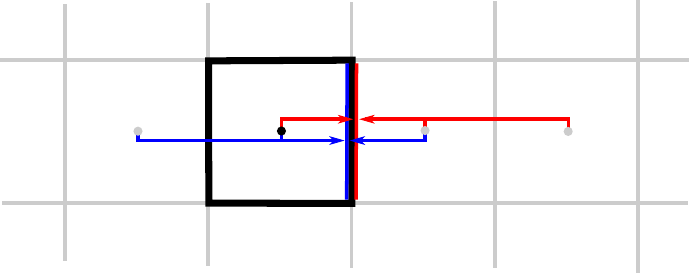}}%
    \put(0.19556496,0.26383189){\color[rgb]{0,0,0}\makebox(0,0)[t]{\lineheight{1.25}\smash{\begin{tabular}[t]{c}$L_1$\end{tabular}}}}%
    \put(0.40721399,0.26383189){\color[rgb]{0,0,0}\makebox(0,0)[t]{\lineheight{1.25}\smash{\begin{tabular}[t]{c}$L_0$\end{tabular}}}}%
    \put(0.61886301,0.26383189){\color[rgb]{0,0,0}\makebox(0,0)[t]{\lineheight{1.25}\smash{\begin{tabular}[t]{c}$R_0$\end{tabular}}}}%
    \put(0.82295314,0.26383189){\color[rgb]{0,0,0}\makebox(0,0)[t]{\lineheight{1.25}\smash{\begin{tabular}[t]{c}$R_1$\end{tabular}}}}%
    \put(0,0){\includegraphics[width=\unitlength,page=2]{sreconstruction.pdf}}%
    \put(0.19567334,0.13700063){\color[rgb]{0,0,0}\makebox(0,0)[t]{\lineheight{1.25}\smash{\begin{tabular}[t]{c}$h_{L_1}$\end{tabular}}}}%
  \end{picture}%
\endgroup%

    \caption{Reconstruction stencil for the structured grid interpolator.}
    \label{sreconstruction}
\end{figure}

Eilmer performs the reconstruction substep differently depending on whether it is operating in structured or unstructured mode. When using a structured grid the cells are arranged into rectangular arrays of hexahedral elements, and each face is aware of its two adjacent cells on each side, labelled $L_1$, $L_0$, $R_0$, $R_1$ in figure \ref{sreconstruction}. This allows the reconstruction algorithm to perform a pair of quadratic interpolations of each variable to the face, one on either side.

The reconstruction of a variable $q$ is then computed as follows, accounting for the possibly nonuniform spacing of the cells $h$,
\begin{equation}
\begin{split}
q_L = q_{L_0} + \alpha_{L} \left[\Delta_{L+} (2 h_{L_0} + h_{L_1}) + \Delta_{L-} h_{R_0} \right] s_L\\
q_R = q_{R_0} - \alpha_{R} \left[\Delta_{R+} h_{L_0} + \Delta_{R-} (2 h_{R_0} + h_{R_1}) \right] s_R
\end{split}
\end{equation}

with the $\Delta$ and $\alpha$ terms:

\begin{equation}
\begin{split}
\Delta_{L-} = \frac{q_{L_0} - q_{L_1}}{1/2(h_{L_0} + h_{L_1})},&~~~ \Delta_{R+} = \frac{q_{R_1} - q_{R_0}}{1/2(h_{R_0} + h_{R_1})}\\
\Delta_{L+} = \Delta_{R-} =&\frac{q_{R_0} - q_{L_0}}{1/2(h_{R_0} + h_{L_0})}\\
\alpha_L = \frac{h_{L_0}/2}{h_{L_1} + 2 h_{L_0} + h_{R_0}},&~~~ \alpha_R = \frac{h_{R_0}/2}{h_{L_0} + 2 h_{R_0} + h_{R_1}}
\end{split}
\end{equation}

\noindent where $s_L$ and $s_R$ are slope limiters using the formulation of \citet{vanalbada_limiter82}, which enforces the conservation of total variation and includes a small constant $\epsilon=1 \times 10^{-12}$ to prevent division by zero.
\begin{equation}
\begin{split}
s_L = \frac{\Delta_{L-} \Delta_{L+} + | \Delta_{L-} \Delta_{L+} | + \epsilon}{\Delta_{L-}^2 + \Delta_{L+}^2 + \epsilon}\\
s_R = \frac{\Delta_{R-} \Delta_{R+} + | \Delta_{R-} \Delta_{R+} | + \epsilon}{\Delta_{R-}^2 + \Delta_{R+}^2 + \epsilon}
\end{split}
\end{equation}

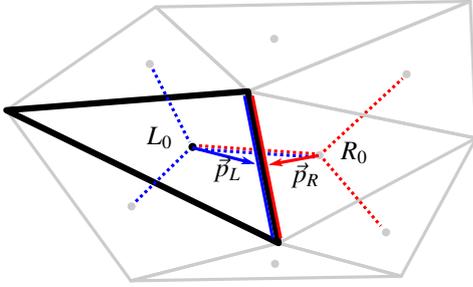
\begin{figure}[h]
    \centering
    \def\svgwidth{0.8\columnwidth}
\begingroup%
  \makeatletter%
  \providecommand\color[2][]{%
    \errmessage{(Inkscape) Color is used for the text in Inkscape, but the package 'color.sty' is not loaded}%
    \renewcommand\color[2][]{}%
  }%
  \providecommand\transparent[1]{%
    \errmessage{(Inkscape) Transparency is used (non-zero) for the text in Inkscape, but the package 'transparent.sty' is not loaded}%
    \renewcommand\transparent[1]{}%
  }%
  \providecommand\rotatebox[2]{#2}%
  \newcommand*\fsize{\dimexpr\f@size pt\relax}%
  \newcommand*\lineheight[1]{\fontsize{\fsize}{#1\fsize}\selectfont}%
  \ifx\svgwidth\undefined%
    \setlength{\unitlength}{56.67744279bp}%
    \ifx\svgscale\undefined%
      \relax%
    \else%
      \setlength{\unitlength}{\unitlength * \real{\svgscale}}%
    \fi%
  \else%
    \setlength{\unitlength}{\svgwidth}%
  \fi%
  \global\let\svgwidth\undefined%
  \global\let\svgscale\undefined%
  \makeatother%
  \begin{picture}(1,0.60030932)%
    \lineheight{1}%
    \setlength\tabcolsep{0pt}%
    \put(0,0){\includegraphics[width=\unitlength,page=1]{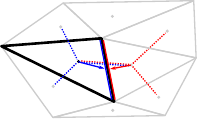}}%
    \put(0.60746023,0.21185055){\color[rgb]{0,0,0}\makebox(0,0)[lt]{\lineheight{1.25}\smash{\begin{tabular}[t]{l}$\vec{p}_R$\end{tabular}}}}%
    \put(0.44371277,0.22605196){\color[rgb]{0,0,0}\makebox(0,0)[lt]{\lineheight{1.25}\smash{\begin{tabular}[t]{l}$\vec{p}_L$\end{tabular}}}}%
    \put(0.3019329,0.29339746){\color[rgb]{0,0,0}\makebox(0,0)[lt]{\lineheight{1.25}\smash{\begin{tabular}[t]{l}$L_0$\end{tabular}}}}%
    \put(0.71191263,0.26149712){\color[rgb]{0,0,0}\makebox(0,0)[lt]{\lineheight{1.25}\smash{\begin{tabular}[t]{l}$R_0$\end{tabular}}}}%
  \end{picture}%
\endgroup%

    \caption{Reconstruction stencil for the unstructured grid interpolator.}
    \label{ureconstruction}
\end{figure}

In unstructured mode (which may also use arrays of hexahedra but is designed to be agnostic as to the grid type), a cell-centred least-squares gradient calculation is performed at each cell (see equations \ref{lsq_extrapolation}-\ref{lsq_gradients}), and the reconstruction is completed with linear extrapolation using the computed gradients. Figure \ref{ureconstruction} shows the cloud of cells used for the gradient calculation in each cell using dotted lines, and the two displacement vectors from the cell centers to the face mid-point $\vec{p}_L$ and $\vec{p}_R$.

The left and right side face values are then reconstructed as follows.
\begin{equation}
\begin{split}
q_L = q_{L_0} + s_L \times \nabla q_L \cdot \vec{p}_L\\
q_R = q_{R_0} + s_R \times \nabla q_R \cdot \vec{p}_R\\
\end{split}
\end{equation}

Once again the $s$ terms are slope limiters, by default the formulation of \citet{venkatakrishnan_limiter95}, which considers the set of all the cells in the local cloud used to generate the gradients $C$ and first computes the minimum and maximum values of the variable.

\begin{equation}
q_{max} = max(q \in C) ~~~;~~~ q_{min} = min(q \in C)
\end{equation}

The limiting function is based on a comparison between two differences in the variable $q$, the first being that predicted by the gradient extrapolated to the face $\Delta_{grad} = \nabla q \cdot \vec{p}$, and the second using either $q_{max}$ or $q_{min}$ for the whole cloud.

\begin{equation}
\Delta_{diff} = \begin{cases}
                  q_{max} - q : \Delta_{grad} > 0\\
                  q_{min} - q : \Delta_{grad} < 0
\end{cases}
\end{equation}

The slope limiter is then computed for each face, and the final value is the minimum of the whole set of faces surrounding the cell.
\begin{equation}
\phi = \frac{ (\Delta_{diff})^2 + 2 \Delta_{diff} \Delta_{grad} + \epsilon}{(\Delta_{diff})^2 + 2 (\Delta_{grad})^2 + \Delta_{diff} \Delta_{grad} + \epsilon}\\
\end{equation}
\begin{equation}
s = min(\phi \in F)
\end{equation}

Once reconstructed, a variety of options are available for computing the inviscid fluxes through a face, ranging from the very dissipative Equilibrium-Flux-Method \cite{macrossan_efm89} and Haenel \cite{hanel_fluxcalc87} flux calculators to the less dissipative AUSMDV \cite{wada_ausm94} scheme, as well as various hybrid options that adaptively switch to a dissipative scheme near a detected shock wave. The flux calculators used in Eilmer are mostly Godunov-type schemes, which consider both sides of the reconstructed interface as a kind of Riemann problem and solve approximately for the state in between to obtain the flux. This approach is attractive because it allows shockwaves and other discontinuities to be captured in the solution with no other special handling, while at the same time producing higher order spatial accuracy in smooth regions of the flow. The different schemes all make different assumptions and approximations in their formulation which leads to situational strengths and weaknesses, and users are encouraged to experiment with the available options to find a flux calculator that works well for their problem.\\

Although the viscous fluxes require less special handling than the inviscid ones, they do require a calculation of the gradients of the primitive variables at each of the cell faces. By default, this task is accomplished using a cell-centred least-squares gradient reconstruction that proceeds as follows.

Consider a first-order extrapolation from the center of a cell to a nearby point, say the centre of one of the faces. If the gradient of a variable $q$ were known, the extrapolation would be computed as:

\begin{equation}
q_f = q_c + \nabla q \cdot \vec{dx}_f
\label{lsq_extrapolation}
\end{equation}

\noindent where $q_c$ is the value of q at cell center, $q_f$ is the value at the face center, and $\vec{dx}_f$ is the distance vector from the former to the latter. In the CFD solver however, the values of $q_f$ and $q_c$ are already known, specifically, $q_f$ is computed as the average of the left and right state reconstructions on each side of the face. This means that the extrapolation will produce a value of $q_f$ with some amount of error, which can be written as:
\begin{equation}
q_f - (q_c + \nabla q \cdot \vec{dx}_f)
\end{equation}

The least-squares gradient calculation proceeds by applying this formula to every face in a given cell, and then solving for the gradient that minimises the square of the error summed across the entire cloud of faces. Formulating this problem as a matrix equation, we begin by presenting the extrapolation step as:

\begin{equation}
\begin{bmatrix}
    w_0 dx_0 & w_0 dy_0 & w_0 dz_0 \\
    w_1 dx_1 & w_1 dy_1 & w_1 dz_1 \\
    w_2 dx_2 & w_2 dy_2 & w_2 dz_2 \\
    w_3 dx_3 & w_3 dy_3 & w_3 dz_3 \\
    w_4 dx_4 & w_4 dy_4 & w_4 dz_4 \\
    w_5 dx_5 & w_5 dy_5 & w_5 dz_5 \\
\end{bmatrix}
\begin{bmatrix}
     \nabla q_x \\
     \nabla q_y \\
     \nabla q_z \\
\end{bmatrix}
=
\begin{bmatrix}
    w_0 (q_0 - q_c) \\
    w_1 (q_1 - q_c) \\
    w_2 (q_2 - q_c) \\
    w_3 (q_3 - q_c) \\
    w_4 (q_4 - q_c) \\
    w_5 (q_5 - q_c) \\
\end{bmatrix}
\end{equation}

\begin{equation*}
\mathbf{X} \cdot \nabla q = \mathbf{dq}
\end{equation*}

\noindent where we have assumed 3D flow with six faces attached to the cell. Eilmer also assigns different weights $w$ to each face based on each one's inverse distance to cell center, effectively weighting nearer faces higher in the error minimisation. The gradient that best minimises the right hand side difference vector $\mathbf{d q}$ is then:

\begin{equation}
\nabla q = (\mathbf{X}^T \mathbf{X})^{-1} \mathbf{X}^T \mathbf{d q}
\label{lsq_gradients}
\end{equation} 

An interesting property of this expression is that the design matrix $\mathbf{X}$ depends only on the cell geometry, which means that the complicated matrix $(\mathbf{X}^T \mathbf{X})^{-1} \mathbf{X}^T$ need only be computed once for each cell during startup, and then at each step of the calculation the gradients $\nabla q$ are evaluated cheaply using the current solutions difference vector. This exact same procedure is used in the unstructured solver to compute the inviscid reconstruction, however a cloud of nearby cells is used, and the weightings $w$ are all set to one.

\begin{figure}[h]
    \centering
    \def\svgwidth{0.8\columnwidth}
\begingroup%
  \makeatletter%
  \providecommand\color[2][]{%
    \errmessage{(Inkscape) Color is used for the text in Inkscape, but the package 'color.sty' is not loaded}%
    \renewcommand\color[2][]{}%
  }%
  \providecommand\transparent[1]{%
    \errmessage{(Inkscape) Transparency is used (non-zero) for the text in Inkscape, but the package 'transparent.sty' is not loaded}%
    \renewcommand\transparent[1]{}%
  }%
  \providecommand\rotatebox[2]{#2}%
  \newcommand*\fsize{\dimexpr\f@size pt\relax}%
  \newcommand*\lineheight[1]{\fontsize{\fsize}{#1\fsize}\selectfont}%
  \ifx\svgwidth\undefined%
    \setlength{\unitlength}{56.48995891bp}%
    \ifx\svgscale\undefined%
      \relax%
    \else%
      \setlength{\unitlength}{\unitlength * \real{\svgscale}}%
    \fi%
  \else%
    \setlength{\unitlength}{\svgwidth}%
  \fi%
  \global\let\svgwidth\undefined%
  \global\let\svgscale\undefined%
  \makeatother%
  \begin{picture}(1,0.33003839)%
    \lineheight{1}%
    \setlength\tabcolsep{0pt}%
    \put(0,0){\includegraphics[width=\unitlength,page=1]{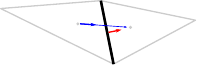}}%
    \put(0.32923744,0.20808113){\color[rgb]{0,0,0}\makebox(0,0)[lt]{\lineheight{1.25}\smash{\begin{tabular}[t]{l}L\end{tabular}}}}%
    \put(0.71062673,0.17119983){\color[rgb]{0,0,0}\makebox(0,0)[lt]{\lineheight{1.25}\smash{\begin{tabular}[t]{l}R\end{tabular}}}}%
    \put(0.5703062,0.21562462){\color[rgb]{0,0,1}\makebox(0,0)[lt]{\lineheight{1.25}\smash{\begin{tabular}[t]{l}$\vec{e}$\end{tabular}}}}%
    \put(0.58036456,0.11950415){\color[rgb]{1,0,0}\makebox(0,0)[lt]{\lineheight{1.25}\smash{\begin{tabular}[t]{l}$\hat{n}$\end{tabular}}}}%
    \put(0.43116215,0.22903628){\color[rgb]{0,0,1}\makebox(0,0)[lt]{\lineheight{1.25}\smash{\begin{tabular}[t]{l}$\hat{e}$\end{tabular}}}}%
  \end{picture}%
\endgroup%

    \caption{Vector definitions for Ref. \cite{haselbacher_discretisation00} face gradient calculation.} 
    \label{face_gradients}
\end{figure}
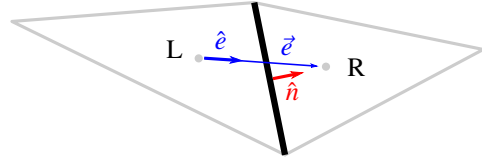

Once the cell-centred gradients are computed, they are then transferred to the faces by an averaging method developed by \citet{haselbacher_discretisation00}, which considers the average gradient from both sides of the face $\nabla w_{av} = \frac{1}{2} (\nabla w_L + \nabla w_R)$ and the alignment between the face normal vector $\hat{n}$ and the vector between the two cell centers $\vec{e}$, as shown in figure \ref{face_gradients}.

The face averaged gradients $\nabla w_f$ are then computed using the following equation.

\begin{equation}
\nabla w_f = \nabla w_{av} - \frac{\left(\nabla w_{av} \cdot \hat{e} - (w_R - w_L)/|\vec{e}| \right)}{\hat{n} \cdot \hat{e}} \hat{n}
\end{equation}

This formula corrects for some amount of cell non-uniformity and is critically important for resolving viscous flows on unstructured grids, but it has also proven to be helpful for structured calculations as well. The extra dissipation appears to help stabilize the calculation especially in the boundary layer, where the gradients are often high and the cell aspect ratios large.\\

Computing the source terms $\mathbf{Q}$ is reasonably straightforward except for the chemical reactions, which are often numerically stiff and require special handling.
When running Eilmer in transient (explicit) mode,
the chemical state of each cell is decoupled from the main flow solver
and advanced using an Ordinary Differential Equation (ODE) solver that
integrates the chemical state through a number of subcycles for each timestep.
For numerically stiff chemical reactions, such as those encountered in combusting flows,
Mott's $\alpha$-quasi-steady-state integrator~\cite{mott_2000} is available.
For other cases, an adaptive Runge-Kutta-Fehlberg scheme \cite{gollan_phd} is used
for integrating the chemical kinetic ODE, which computes both a fifth-order accurate and sixth-order accurate solution of the ODE and adjusts the substep size to keep the difference between the two within a tolerance. When running in implicit (steady-state) mode,
no subcycling is performed and the source terms for the chemistry are calculated directly,
relying on the inherent stability of the implicit update scheme to handle the chemical stiffness.\\

The final substep of the simulation main loop is to compute the rate of the change of the conserved variables (sometimes called the residual $\mathbf{R}$) using equation \ref{semidiscrete} and update the conserved variables using a time-stepping scheme.
\begin{equation}
\frac{\partial \mathbf{U}}{\partial t} = \mathbf{R}
\label{residual}
\end{equation}

Eilmer has a number of options for these schemes, including a family of explicit integrators based on the the Runge-Kutta approach, where the residual is computed multiple times over a single step and the results blended together to achieve higher-order time accuracy. For example, a two-stage predictor corrector explicit scheme computes the change in conserved variable from timestep $n-1$ to $n$ as follows.\\

\begin{equation*}
\mathbf{U}^{n+1}_{pred} = \mathbf{U}^{n} + \mathbf{R}(\mathbf{U}^n) dt
\end{equation*}

\begin{equation}
\mathbf{U}^{n+1} = \mathbf{U}^{n} + \left(\frac{1}{2} \mathbf{R}(\mathbf{U}^{n}) + \frac{1}{2} \mathbf{R}(\mathbf{U}^{n+1}_{pred})\right) dt
\end{equation}

For some problems, it is possible to gain significant improvements in run time by using implicit time advancement schemes that are stable at large timesteps. Eilmer contains a point-implicit update scheme which is based on a Backward-Euler formulation where the change in conserved quantities $\Delta \mathbf{U}$ is computed using the unknown residual at time $n+1$. This residual may be unknown, but it can be approximated using a first-order linearisation that couples the updates at every point of the flow into a large linear system.

\begin{equation}
\frac{\Delta \mathbf{U}}{\Delta t} = \mathbf{R}^{n+1} = \mathbf{R}^{n} + {\frac{\partial \mathbf{R}}{\partial \mathbf{U}}}^n \Delta \mathbf{U}^n
\end{equation}

Introducing the identity matrix $\mathbf{I}$, this can be rearranged into a large sparse matrix problem as follows.
\begin{equation}
\left[ \frac{1}{\Delta t} \mathbf{I} + \frac{\partial \mathbf{R}}{\partial \mathbf{U}}^n \right] \Delta \mathbf{U}^n = \mathbf{R}^{n}
\label{implicit_matrix_equation}
\end{equation}

Equation \ref{implicit_matrix_equation} introduces the Jacobian matrix $\partial \mathbf{R}/\partial \mathbf{U}$, a measure of the sensitivity of the flow time derivatives to the conserved variables. To evaluate the Jacobian, Eilmer uses a complex-step finite differencing approach, which proceeds by perturbing each component of $\mathbf{U}$ by a small amount in the imaginary plane and comparing the resulting difference in $\mathbf{R}$. 
\begin{equation}
\frac{\partial \mathbf{R}_i }{\partial \mathbf{U}_j} = \frac{Im(R_i(U_j + i h))}{h}
\end{equation}

The complex number differentiation is implemented using the D language's templating capability, and has proven to be a significant improvement over using real-valued finite differencing, which often experiences difficulties related to the choice of perturbation size, or analytic differentiation of the code, which would be a major maintenance and development burden.

An exact solution to the matrix problem in equation \ref{implicit_matrix_equation} would be expensive to compute and is not generally necessary, since for steady-state problems we are only interested in reducing $\mathbf{R}$ with each iteration. The point-implicit calculation thus approximates by setting all of the off-diagonal terms in equation \ref{implicit_matrix_equation}'s block matrix to zero, effectively uncoupling the cells from each other and turning the update problem into an isolated matrix problem for each cell that can be solved one-at-a-time. This produces a time advancement method with first-order accuracy that is stable at large timesteps, allowing the solver to rapidly approach a steady-state solution that would otherwise take an impractically large number of explicit steps. Eilmer also has an experimental solver specifically designed for steady flows, in which equation \ref{implicit_matrix_equation} is solved for $\mathbf{R}=0$ using a series of pseudo-timesteps by a Jacobian-Free Newton-Krylov method.

\section{Development Processes}
\label{sec:development}
\begin{quote}
\emph{Software is as important to modern scientific research as telescopes and test tubes.}
\end{quote}
{\raggedleft Wilson et al. (2014) \cite{wilson_scicomp14}\\ \par}

\vspace{5mm}
The earliest Eilmer code can be traced back to the early 1990's, when the era of modern CFD was inaugurated by the invention of parallel hardware and advances in applied mathematics like high-resolution flux calculators. In the years since, the code has evolved through multiple languages, rewrites, and paradigm shifts, and we have accumulated some lessons learned that may be of interest to readers of this article. This section describes the behind the scenes detail that goes into developing the code; the programming practices, verification and validation exercises, and testing protocols that we use to write new code and ensure it works properly. It follows in the spirit of articles such as \citet{wilson_scicomp14} that call for scientific computing to be more responsibly conducted and better scrutinised, noting as they do that the risk of mistakes in our field is high and the consequences potentially severe. We begin with the programming framework and discuss the reasons for our somewhat unusual choice of language and setup, then describe the various forms of quality control and testing the code is subjected to. 
\pagebreak

\subsection{The D Programming Language}
Eilmer is written in D: A high-level, multi-paradigm, statically-typed, compiled programming language intended to be a modernised revision of C++ with the same speed and flexibility. Though D maintains a familiar C-like syntax, it features substantially redesigned high-level features such as objects, templates, and automatic memory management, based on lessons learned from C++ and other popular languages. Specific features of D that have helped us include string mixins for automatically generated inline code, improved templating syntax with much clearer error messages when things go wrong, a module-based architecture that removes the need for header files, and improved object-orientated syntax with property functions that eliminates the need for verbose getters and setters on all of an object's public-facing attributes.

Eilmer itself was written in C++ during the 2006-2015 era of the code's lifetime, but the 2016-present implementation has greatly benefited from D's many improvements in terms of overall line count, complexity, and programmer-friendliness. Our project in particular is maintained part-time by a small team of researchers, and D's helpful error messages and reduced boilerplate has allowed us to make much better use of our limited development time. The code is also used by many students and faculty who are not professional programmers, and the shallower learning curve compared to C++ ensures they are able to make genuine contributions in their teaching assignments and thesis projects.

\begin{figure}[h]
    \small
    \centering
    \def\svgwidth{1.0\columnwidth}
\begingroup%
  \makeatletter%
  \providecommand\color[2][]{%
    \errmessage{(Inkscape) Color is used for the text in Inkscape, but the package 'color.sty' is not loaded}%
    \renewcommand\color[2][]{}%
  }%
  \providecommand\transparent[1]{%
    \errmessage{(Inkscape) Transparency is used (non-zero) for the text in Inkscape, but the package 'transparent.sty' is not loaded}%
    \renewcommand\transparent[1]{}%
  }%
  \providecommand\rotatebox[2]{#2}%
  \newcommand*\fsize{\dimexpr\f@size pt\relax}%
  \newcommand*\lineheight[1]{\fontsize{\fsize}{#1\fsize}\selectfont}%
  \ifx\svgwidth\undefined%
    \setlength{\unitlength}{432bp}%
    \ifx\svgscale\undefined%
      \relax%
    \else%
      \setlength{\unitlength}{\unitlength * \real{\svgscale}}%
    \fi%
  \else%
    \setlength{\unitlength}{\svgwidth}%
  \fi%
  \global\let\svgwidth\undefined%
  \global\let\svgscale\undefined%
  \makeatother%
  \begin{picture}(1,0.34468163)%
    \lineheight{1}%
    \setlength\tabcolsep{0pt}%
    \put(0,0){\includegraphics[width=\unitlength,page=1]{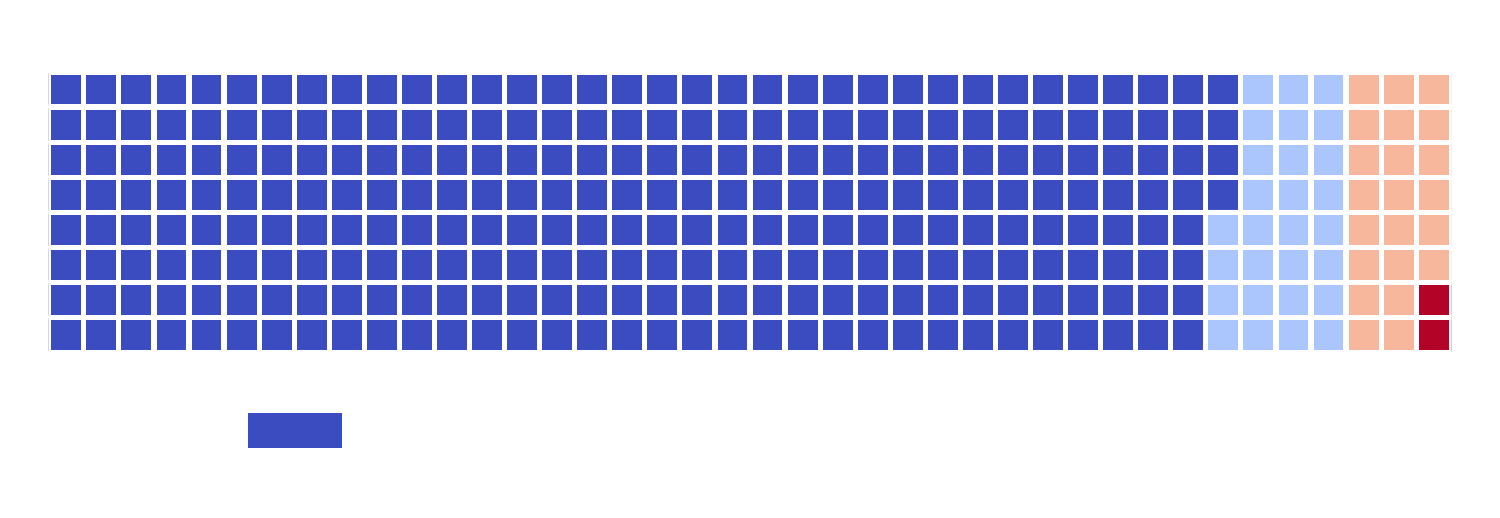}}%
    \put(0.2506696,0.0469621){\makebox(0,0)[lt]{\lineheight{1.25}\smash{\begin{tabular}[t]{l}D - 83.7\%\end{tabular}}}}%
    \put(0,0){\includegraphics[width=\unitlength,page=2]{lines_of_code.pdf}}%
    \put(0.2506696,0.00279179){\makebox(0,0)[lt]{\lineheight{1.25}\smash{\begin{tabular}[t]{l}Lua - 6.9\%\end{tabular}}}}%
    \put(0,0){\includegraphics[width=\unitlength,page=3]{lines_of_code.pdf}}%
    \put(0.56596004,0.0469621){\makebox(0,0)[lt]{\lineheight{1.25}\smash{\begin{tabular}[t]{l}Python - 8.7\%\end{tabular}}}}%
    \put(0,0){\includegraphics[width=\unitlength,page=4]{lines_of_code.pdf}}%
    \put(0.56596004,0.00279179){\makebox(0,0)[lt]{\lineheight{1.25}\smash{\begin{tabular}[t]{l}Other - 0.7\%\end{tabular}}}}%
  \end{picture}%
\endgroup%

    \caption{Division of languages within Eilmer as of April 2022.}
    \label{languages}
\end{figure}

Figure \ref{languages} shows the present division of D and other languages present in the repository as a waffle plot, where the area or number of squares corresponds to the proportion of each language.

Like many programs, Eilmer uses Lua as an internal scripting language for configuration and run-time customisation. This is facilitated by D's C library bindings, which are used to pass data in and out of a Lua interpreter that is co-compiled and linked into the main executable. The source code for this interpreter is a small amount of lightweight C code that is carried around in our repository, part of a general philosophy that tries to minimise external dependencies that the user has to obtain before compiling the code. This philosophy is a lesson learned from the previous incarnation of the code, which suffered from a degree of churn and rot in the many open-source libraries and toolkits it relied on. The current implementation requires only on a handful of standard Linux development libraries, a D compiler, and OpenMPI for parallelisation, with a few optional extra features available if the foreign-function libraries for Python and Ruby are installed. In this manner it is compatible with a wide range of High-Performance Computing (HPC) systems and should remain so even as hardware and software trends shift in the future.

\subsection{Verification and Quality Assurance}
\label{verification}
Modern scientific computing distinguishes two ways of vetting that software is fit for purpose. Verification (\emph{``are we solving the equations right''}) involves checking that the program is correctly achieving the intentions of its programmers. An incorrectly spelled variable, an unintended logical comparison, or a mistake in a mathematical formula are all examples of problems that are in the realm of verification. The second activity, Validation (\emph{``are we solving the right equations''}) is the subject of the section \ref{validation}.

The first technique applied to the verification of Eilmer will be familiar to software engineers: unit tests. A core principle of good software design is to break a program up into encapsulated modules, each with a single clearly defined responsibility. To ensure that the smallest subcomponents of the code are performing correctly, each can be tested in isolation by being given a small test problem and comparing the output to a known good value. 

Extensive unit tests are available for Eilmer's subcomponents, including the modules for handling gas behaviour, chemical kinetics, geometry, and numerical methods/linear algebra. For example, the gas tests can be invoked using:

\begin{verbatim}
$ cd src/gas
$ make test
\end{verbatim}

This will compile a small executable for each of the relevant gas models and use them to compute a set of sample problems, such as the internal energy of a mixture given a pressure, temperature, and composition. The output of each calculation is then compared to a known good value, producing a summary of the outcomes and a success message if all is well.\\

\begin{figure}[h]
    \tiny
    \def\svgwidth{1.0\columnwidth}
    \input{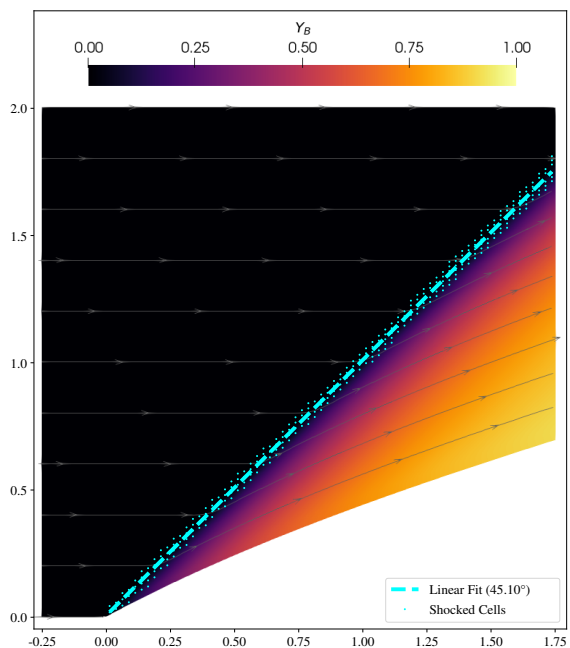}
    \caption{Oblique detonation wave, mass fraction of species B.}
    \label{odw_yB}
\end{figure}

Another technique of verification used within Eilmer is to compare an entire simulation against an exact solution computed using an analytic technique. Although the equations of fluid mechanics are, in general, too complex to be solved by these techniques, for some special cases exact analytical solutions are available. One example of this procedure applied to Eilmer is shown in figure \ref{odw_yB}, a solution of the detonation wave problem proposed by \citet{powers_detwave06} in a paper calling for more widespread and rigorous verification of numerical codes.

The problem consists of a curved ramp that drives an oblique detonation wave through the oncoming flow, sustained by a simplified one-step chemical reaction that activates once a temperature threshold has been reached. The reaction rates and the curved ramp surface are designed to ensure that the angle of wave is fixed at 45$^\circ$, and exact solutions are available for postshock flow conditions along streamlines through the shocked gas. Figure \ref{odw_yB} shows the solver's computed solution, with a regression line fitted to the cells where the shock is detected. Even on a relatively coarse grid this agrees well with the analytic solution, and further refinements show good results for the postshock flowfield as well, evidence that the reaction rates and the chemistry/flow solver coupling has been implemented correctly.\\

Computational physics calculations can also be verified using the Method of Manufactured Solutions. This technique takes advantage of an interesting feature of the partial differential equations used in continuum mechanics codes, which typically have the following form.

\begin{equation}
\frac{\partial \mathbf{U}}{\partial t} +
\frac{\partial \mathbf{F}\left(\mathbf{U}\right)}{\partial x} +
\frac{\partial}{\partial x} \frac{\partial \mathbf{V}\left(\mathbf{U}\right)}{\partial x} = 0
\end{equation}

In general, any given $\mathbf{U}$ will not be a solution to these equations, and putting it through the numerical machinery for calculating the derivatives would give a non-zero remainder on the right hand side. The essential insight of the Method of Manufactured solutions is to add an artificial source term to the equations that precisely matches this remainder.

\begin{equation}
\frac{\partial \mathbf{U}}{\partial t} +
\frac{\partial \mathbf{F}\left(\mathbf{U}\right)}{\partial x} +
\frac{\partial}{\partial x} \frac{\partial \mathbf{V}\left(\mathbf{U}\right)}{\partial x} =
\frac{\partial \mathbf{F}\left(\mathbf{U}_{mms}\right)}{\partial x} +
\frac{\partial}{\partial x} \frac{\partial \mathbf{V}\left(\mathbf{U}_{mms}\right)}{\partial x}
\end{equation}

By choosing an arbitrary solution $\mathbf{U}$ that is analytically differentiable, the source term on the right hand side can be evaluated exactly. One can then run the numerical solver with this manufactured source term in every cell, which acts to drive the numerical solution toward $\mathbf{U}_{mms}$, assuming that the numerical machinery that approximates the spatial derivatives has been implemented correctly. The discrepancy between the final value arrived at by the solver and the manufactured solution is then an estimate of the discretisation error in the numerical algorithm, which should be small if all is well. 

Eilmer has several manufactured solution test cases included in its examples directory and others that are published in various places, which provide verification coverage across both structured \cite{gollan_eilmer13}, and unstructured \cite{damm_phd} grids, as well as solid/fluid conjugate heat transfer flows\cite{eilmer3_mms_2016}. A new example of a manufactured solution, this one for a viscous flow with the Spalart-Allmaras turbulence model, is shown in figure \ref{cube}. (The manufactured solution itself is documented in Appendix 1.)

\begin{figure}[h]
    \tiny
    \def\svgwidth{1.0\columnwidth}
    \input{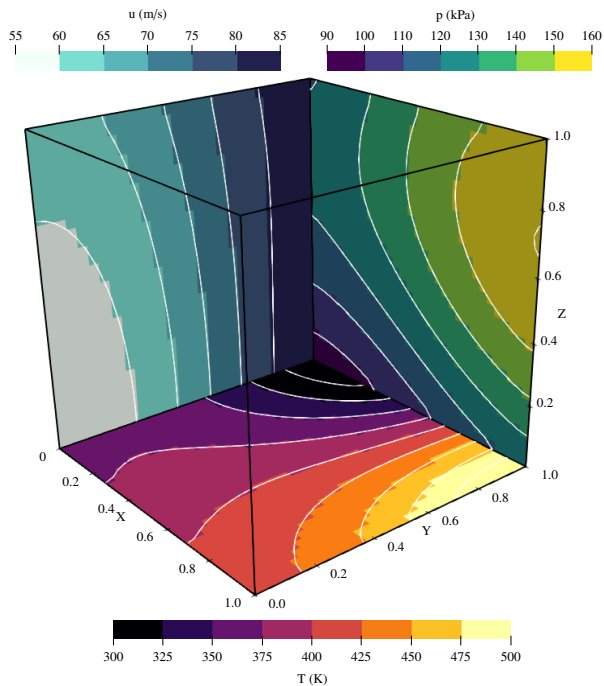}
    \caption{Colour maps of numerical solution to MMS test case. Exact solution shown in white contour lines.}
    \label{cube}
\end{figure}

\begin{figure}[h]
    \scriptsize
    \def\svgwidth{1.0\columnwidth}
    \input{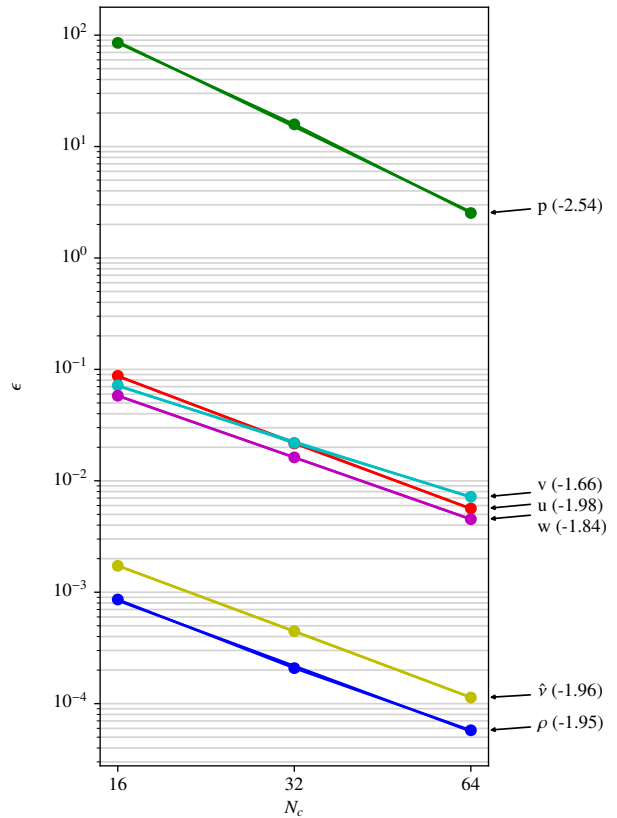}
    \caption{Error norms and orders of accuracy from 3D MMS cases, with 16$^3$, 32$^3$, and 64$^3$ cells}
    \label{norms}
\end{figure}

Merely converging the the manufactured solution is a necessary but not sufficient condition for verification. A key characteristic of the numerical methods used in high-resolution CFD solvers is that they are higher order: the discretisation error should not merely reduce as the grid is refined, but the reduction will be nonlinear, generally second order $O(dx^{2})$. This property can be checked by computing the total error of multiple MMS cases with different levels of refinement, and plotting the results on a logarithmic scale as in figure \ref{norms}. The test shows the expected nonlinear drop, with the slope of each line indicated by an annotation for each variable.

\begin{figure*}[t]
    \centering
    \def\svgwidth{0.8\textwidth}
    \input{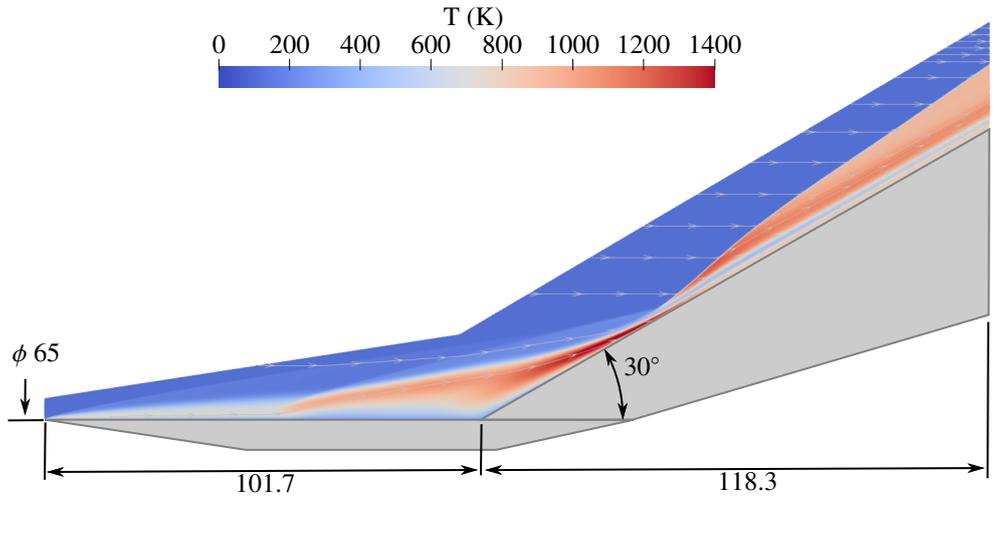}
    \caption{Temperature colour map of hollow cylinder flare solution. Dimensions in mm.}
    \label{cyl}
\end{figure*}

\subsection{Validation and Reality Checks}
\label{validation}
Verification and Validation are the oft-confused twin pillars of responsible scientific computing. The previous section has dealt with Verification: Are we solving the equations correctly? This subsection takes up Validation: Are we solving the correct equations? 

Validation involves checking a program's correspondence with the aspects of reality it is intended to simulate. Comparing numerical predictions against experimental data is the gold-standard activity, but other valuable work includes checking the validity of assumptions in the formulation and establishing boundaries where missing physics models are required. In contrast to Verification, Validation is not so much the activity of a single person but a shared responsibility of the entire research community, as it relates to the shared pool of physical models we have developed and our knowledge of their limitations.  

In this regard Eilmer benefits from a long history of collaboration with experimental hypersonics. The ancestry of the code can be traced back to cns4u, a single block Navier-Stokes integrator written by Peter Jacobs in the early 90's for simulating hypersonic wind tunnels. Testing against measurements taken in an expansion tube revealed that the program could predict the large scale wave processes reasonably well, as measured by pressure sensors on the inside of the tunnel walls and a rake of pitot-pressure sensors in the freestream \cite{jacobs_expansiontubes94}. Simulations of wind tunnel operation continue today using the modern incarnation of the code, for example in \citet{jewell_quietstartup17} who performed transient 2D calculations of the AFRL Mach 6 Ludwieg tube, and \citet{gildfind_x3cfd18} simulations of the large X3 expansion tube. Both authors report good agreement with pressure measurements taken inside the respective facilities, though the expansion tunnel behaviour is complex and not perfectly captured for various reasons of approximation.

\begin{figure}[h]
    \tiny
    \def\svgwidth{1.0\columnwidth}
    \input{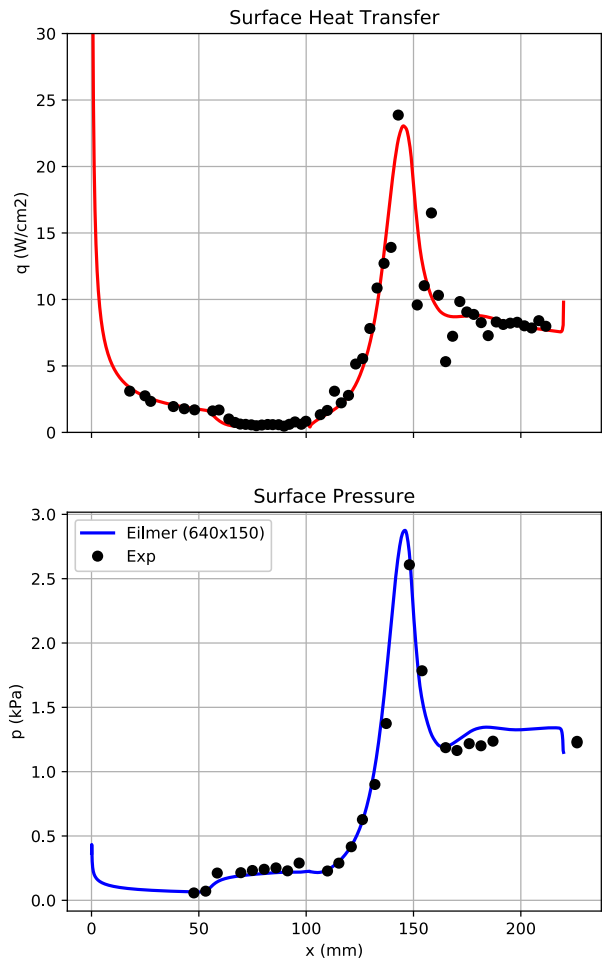}
    \caption{Surface pressure and heat transfer results from hollow cylinder flare solution. Experiments by Ref. \cite{holden_cubrc03}}
    \label{cyl_results}
\end{figure}

A common use for hypersonic wind tunnels is to generate experimental data solely for the purpose of validating CFD. Basic shapes such as ramps, cavities, and spheres are typically used for this purpose, outfitted with pressure and heat transfer sensors that measure the state of the flow touching the model. An example is the hollow flared-cylinder model tested by \citet{holden_cubrc03} in the CUBRC shock tunnel facilities, part of a large library of experimental results intended to be used for validation studies. The flow over the outside of the cylinder features a separation and a viscous shock interaction, shown in figure \ref{cyl}, as computed by an Eilmer simulation that is available in the examples directory online. The predicted heat load and pressure on the walls compare quite closely to the experimental measurements (see figure \ref{cyl_results}), and are in line with the predictions of other hypersonics codes reported in \citet{maclean_validation04}.

Similar exercises have been reported in \citet{sun_conecfd05}, in which multiple codes (including Eilmer) are compared on a cone-shock interaction problem; \citet{park_cylinder16} who matched pressure and heat transfer results on a high-enthalpy cylinder flow; and \citet{damm_phd}, who documents four different validation tests including a double cone, shock interaction, and a turbulent boundary layer, all non-reacting. An additional reacting flow exercise can be found in \citet{hoste_cfd17}. In general, these exercises generate reasonably good agreement to pressure measurements, though heat transfer is sometimes harder to match due to the larger margins of error encountered in the measurements. This problem has been underlined by multi-code validation exercises such as \citet{candler_cfd17}, where consistent problems matching certain kinds of experimental data have been found, especially in high enthalpy flows. Interestingly, \citet{ray_inflowuncertainties20} have suggested that the major cause of these problems could be that the inflow conditions given to the CFD solvers are often inaccurate, due to the large number of approximations made in the facility flow models that generate the predictions of what the conditions are. Better numerical simulations of hypersonic facilities (such as those described in an earlier paragraph) are an obvious place to look for answers to this problem.\\

More direct measurements of a hypersonic flow can be obtained using optical techniques such as Schlieren photography. A handful of experiments (such as \citet{lobb_sphere62} and \citet{nonaka_shockstandoff00}) have used this technique to visualise the shockwave created by a blunt hypersonic object as it flies through a ballistic range, a useful arrangement that avoids the problem of reconstructing the test flow from a wind tunnel. The distance between the object's leading edge and its bow shock (as detected by the Schlieren) is called the shock standoff distance, which can be used for code validation if the precise shape of the shock can be reconstructed from the image using a fitting technique. 

A collection of these exercises, performed using Eilmer, is reported in \citet{gollan_shockstandoff12}, with more detailed analysis undertaken by \citet{zander_shockstandoff14} for higher enthalpy flows. In general, the code performed well, predicting the shock shape to within the range of experimental error, as long as thermal and chemical nonequilibrium are accounted for. Though these results affirm the presence of nonequilibrium effects caused by high temperatures, the data are currently too noisy to distinguish which of the many reaction schemes and relaxation models should be preferred.\\

Further insight into questions like this can be gained using finer grained optical diagnostics, such as emission spectroscopy, to interrogate the bright light emitted by a hot hypersonic flow and separate it into frequencies. In theory, one can even take the temperatures and composition computed by a CFD simulation, input them into a radiation program that accounts for all the relevant lines, transitions, and atomic processes, and compare the resulting artificial spectrum to a real one measured in an experiment. In practice, these calculations are fiendishly complicated and often have discrepancies that are difficult to diagnose, given the many steps between the input and output. Studies such as \citet{fahy_hayabusa21} have used Eilmer and the PARADE radiation database to compute the radiation from the 2010 Hayabusa probe reentry, finding reasonably good agreement with measurements in the near-infrared frequencies (mostly oxygen and nitrogen lines), but poor agreement in the ultraviolet range. Similar work by \citet{banerji_venusentry18} considered a Venus-like CO$_2$-N$_2$ atmosphere, using Eilmer and the NEQAIR radiation database to produce reasonable predictions of the visible wavelength range, although a number of lines generated by the complex C$_2$ and CN radiators were missing, with particular trouble again in the ultraviolet.

More interestingly (at least for validation purposes), it is possible to take spectral information measured in a real flow and work backwards to compute other properties such as temperature or composition. One example of this technique was developed by \citet{potter_co2N2radiation08}, who performed least-squares fitting of an artificial spectrum (generated by the \emph{Photaura} radiation calculator) to a real spectrum generated by a CO$_2$-N$_2$ flow over a cylinder in an expansion tube. The fitting produced a temperature and composition that best matched the measured radiation, which could be compared to an Eilmer CFD result. Although the error bars of the resulting temperatures were fairly large ($\approx 10-20\%$), the results were good enough to make some judgements about some of the different CFD models present in the literature. This same technique was later used by \citet{banerji_earthentry18}, this time in an Earth-like atmosphere, to diagnose an abundance of N2$^+$ that was causing overprediction of radiation in the ultraviolet range, and by \citet{gu_co2radiation21} to show that the existing energy relaxation rates of CO$_2$ are far too slow in an expanding flow, such as behind the shoulder of a reentry capsule. Even the spectral lines themselves can be used for code validation. \citet{liu_saturnentry22} measured the width of Hydrogen lines in a simulated gas-giant entry and used the amount of Stark broadening to estimate the concentration of free electrons in the shock layer. These measurements confirmed some suspected problems in the ionisation rate models that are available in the literature, and the authors were able to modify an existing model to produce an Eilmer calculation that showed good agreement with their measurements.\\


\subsection{Continuous Integration Testing}
Any quality control process is only as good as the version of the code it makes use of. Eilmer is a large, constantly evolving codebase with many contributions from collaborators, and we have developed a continuous testing program to keep the code in working order as we add and change things. The core of this process is a set of test simulations (fifty-six at the time of this writing) that can be automatically run and post-processed by a set of scripts written in the Ruby scripting language. The simulations are typically coarsened or scaled down versions of the tests discussed in the prior sections, including simple validation exercises, MMS verification tests, flows with analytic solutions, and a handful of other building block simulations that exercise major features of the code.

At the end of each test the associated script extracts a handful of diagnostic numbers and compares them to a known good outcome; this may be the number of iterations taken by the solver, or the residuals at the final step, or the drag on a flat plate, or the root-mean-square error between the numerical solution and its analytical counterpart. A too-large difference in any diagnostic is reported as a failed test, which typically indicates some kind of bug or regression in the code from the prior commit. 

\begin{figure}[h]
    \scriptsize
    \def\svgwidth{1.0\columnwidth}
    \input{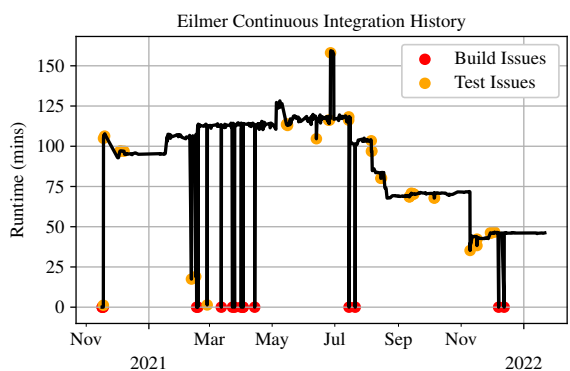}
    \caption{Runtime history of Eilmer continuous testing system.}
    \label{cilog}
\end{figure}

Building a test suite with good coverage is helpful for quality control reasons, but we have found the biggest advantage is that we can do \emph{continuous} testing, using an automated system that tests every published commit and reports via email if something is wrong. This system was implemented in the last quarter of 2020 and has been monitoring our repository ever since. A record of its activity through early 2022 is shown in figure \ref{cilog}.

The execution time of the entire suite has dropped from over an hour to about forty-five minutes throughout 2021 because of optimisations and changes to the roster of tests, and the period includes 509 successful runs, 16 issues with the build process, and 40 failures of the automated tests for the entire period. Many of these issues arose from compiler changes, platform specific glitches, and other obscure problems that are hard for an isolated developer to detect on their own, and it is fair to say we were surprised at how helpful the automated testing system turned out to be. Continuous testing can be easily implemented in a Python script and then run as a cron job using a server or virtual cloud machine, and we highly recommend it to the authors of other scientific and research codes.

\FloatBarrier
\begin{table*}[b]
\centering
\begin{tabular}[h]{rccccccc}
    n Processors &  48 & 96 & 144 & 288 & 576 & 768 & 960 \\
    \hline
    time/iter    &  1.00 & 0.506 & 0.343 & 0.191 & 0.109 & 0.076 & 0.065 \\
    speed-up      &  48.0 & 95.0 & 140.1 & 250.6 & 439.7 & 632.16 & 741.6 \\
\end{tabular}
\caption{Results of parallel scaling tests using 3D blunt cone simulation}
\label{oscar_test_results}
\end{table*}

\section{Parallel Scaling}
\label{sec:scaling}
\begin{figure}[h]
    \centering
    \includegraphics[width=0.99\columnwidth]{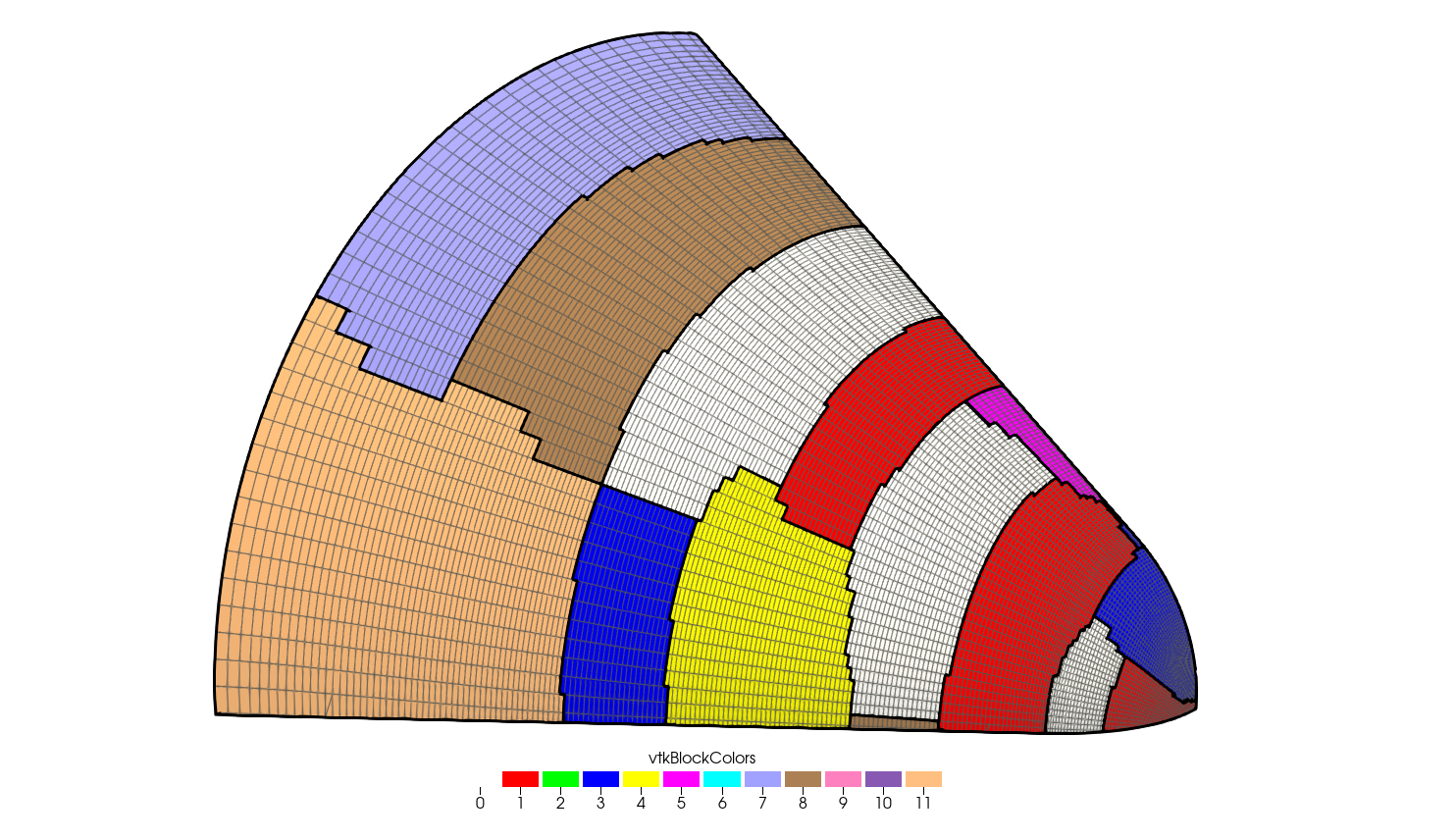}
    \caption{Block decomposition for a viscous blunt cone mesh using METIS, nblocks=48}
    \label{cell_decomposition}
\end{figure}

Fluid dynamics problems can require large amounts of computation time. Like most modern scientific codes, Eilmer can mitigate this problem by breaking a simulation up into a number of independent pieces and executing them in parallel, using multi-core CPUs or groups of them joined by a high-speed network. The core parallelisation strategy we use is \emph{block decomposition}, in which the fluid domain to be simulated is divided into a jigsaw puzzle of non-overlapping blocks, either manually by specifying the split points in a structured grid, or automatically using the external graph partitioning tool METIS \cite{metis}. Figure \ref{cell_decomposition} is an example of a 3D blunt cone partitioned using the automated procedure. 

Each block consists of cell elements that contain an average state of the fluid variables $\mathbf{U}$ throughout a small region of space. The change in state over a small step in time $\Delta \mathbf{U}$ is computed using the fluxes of each variable $\mathbf{F}$, which depend on both the cell in question and also its neighbour, due to the reconstruction process discussed in section \ref{numerical_discretisation}. This spatial dependency makes parallelising the solver a nontrivial problem at the boundaries where one block meets another.

The main way that this problem is solved is by surrounding each block in a layer of virtual or ``ghost'' cells that are not part of the actual simulation, and instead are simply filled with flow that matches the conditions in each neighbouring block. The required data is passed back and forth between the different processors working on each block, using standardised MPI (Message Passing Interface) library functions, along with Reduce and Barrier operations to synchronise control flow between the different processors. In this way, each processor marches independently through the same algorithm but working only on its designated region of the flow.

In this section, we present the results of some scaling experiments to benchmark the code's parallel performance and ensure the implementation is scaling efficiently to the large number of cores that are sometimes required for timely fluid dynamics research. Parallel efficiency can be assessed by using a simple model of task scheduling where an algorithm is split into a Serial fraction $s$ and a Parallel fraction $p$, where $p + s = 1$. The expected time taken $T_n$ for a task running on $n$ processes is:

\begin{equation}
T_n = s T_1 + \frac{p}{n} T_1
\label{Tn}
\end{equation}

The parameters $s$ and $p$ are not actually constants, but they can be estimated for a typical simulation that is representative of the large parallel calculations our users perform on supercomputers. This section presents the results of timing tests on such a case, the 5$^\circ$ blunt cone depicted in figure \ref{cell_decomposition}, which uses a 2 million cell 3D viscous mesh immersed in a thermally perfect air flow at Mach 3. The exact same simulation is performed multiple times on different numbers of processors, and the average time per iteration and estimated speedup is shown in table \ref{oscar_test_results}.

\begin{figure}[h]
    \scriptsize
    \centering
    \def\svgwidth{0.99\columnwidth}
    \input{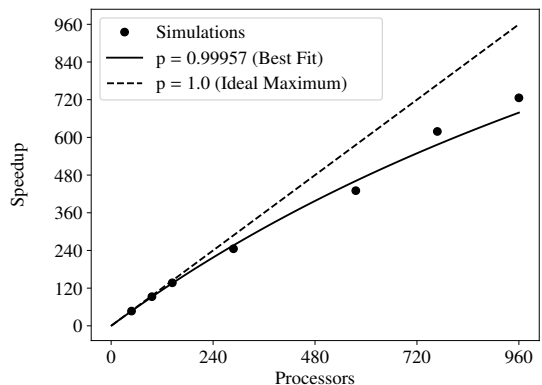}
    \caption{Parallel scaling performance with $p=0.99957$ on blunt cone problem.}
    \label{speedup_cpus}
\end{figure}

In a perfectly parallel calculation the speedup would exactly follow the number of processors, but the table shows speed-up degrading as more and more cores are added, an unavoidable reality of each one having to duplicate the serial fraction $s$ of the algorithm. The above data can be used to curve-fit values of $s$ and $p$ from equation \ref{Tn}, resulting in a parallel efficiency of $p=0.99957$ and the estimated performance shown in figure \ref{speedup_cpus}.

These tests confirm that the parallelisation is scaling efficiently up to $\approx 1000$ processors, \emph{for this particular problem}. Note however that this will not be true in general. Specifically, the efficiency $p$ is a function of the number of cells in the mesh, and a smaller simulation than the 2 million cell test problem described here will be expected to have a lower $p$ value and much faster degradation in the speed-up as the number of cores is increased. Users are encouraged to perform a pair of scaling simulations like in the above example for their own simulations whenever parallel efficiency is important, such as when using a shared cluster machine with metered usage.

\FloatBarrier
\section{Example Applications}
\label{sec:applications}
This section showcases a handful of representative simulations that cover most of the common applications that Eilmer has been used for in the past. They range from pure research in fluid dynamics to applied calculations of heat transfer and pressure in an engineering problem, with aerodynamic shape optimisation and hypersonic facility design somewhere in between. The final example is slightly different --- it consists of a simple blunt body flow calculation performed by our students each year in the University of Queensland's computational fluid dynamics course. Taken together, the examples give a broad outline of what Eilmer is capable of, though the range future applications is limited only by the imagination and, of course, by the available compute time.

\subsection{Minimal Working Example}
The very first simulation that most users of Eilmer will run is the classic sharp-nosed 20$^\circ$ cone, a basic test case that consists of a Mach 1.5 freestream and an oblique shockwave in ideal, inviscid, air. Figure \ref{cone20_config} shows the entire Lua configuration file required to set up the simulation, including building the grid, setting the gas model and boundary conditions, and configuring the solver. 
\begin{figure}[h]
\scriptsize
\begin{lstlisting}
-- 1. Flow domain and grids, dimensions in metres.
config.axisymmetric = true
--
a0 = {x=0.0, y=0.0};     a1 = {x=0.0, y=1.0}
b0 = {x=0.2, y=0.0};     b1 = {x=0.2, y=1.0}
c0 = {x=1.0, y=0.29118}; c1 = {x=1.0, y=1.0}
--
quad0 = CoonsPatch:new{p00=a0, p10=b0, p11=b1, p01=a1}
quad1 = AOPatch:new{p00=b0, p10=c0, p11=c1, p01=b1}
--
grid0=StructuredGrid:new{psurface=quad0,niv=11,njv=41}
grid1=StructuredGrid:new{psurface=quad1,niv=31,njv=41}
--
-- 2. Gas model and flow states.  SI units.
setGasModel('ideal-air-gas-model.lua')
initial = FlowState:new{p=5955.0, T=304.0} -- Pa K
inflow = FlowState:new{p=95.84e3, T=1103.0,
                       velx=1000.0} -- Pa K m/s
--
-- 3. Fluid blocks, initial flow, boundary conditions.
blk0=FluidBlock:new{grid=grid0,initialState=inflow}
blk1=FluidBlock:new{grid=grid1,initialState=initial}
identifyBlockConnections()
blk0.bcList['west']=InFlowBC_Supersonic:new{
                                    flowState=inflow}
blk1.bcList['east']=OutFlowBC_Simple:new{}
--
-- 4. Simulation parameters.
config.max_time = 5.0e-3  -- seconds
config.max_step = 3000
config.dt_plot = 1.5e-3
\end{lstlisting}
\caption{Sharp-nosed 20$^\circ$ cone configuration script}
\label{cone20_config}
\end{figure}

Before running the simulation, this script is passed to a preprocessor that produces Eilmer-native grid and flow solution files, as well as a verbose machine-readable configuration file (in JSON format) with all of the available solver settings, including default values for any that are not specified in the input script. The actual simulation then proceeds using these preprocessed files and not the user-facing Lua input script, eventually producing the flowfield shown below.
\begin{figure}[h]
    \small
    \def\svgwidth{1.0\columnwidth}
\begingroup%
  \makeatletter%
  \providecommand\color[2][]{%
    \errmessage{(Inkscape) Color is used for the text in Inkscape, but the package 'color.sty' is not loaded}%
    \renewcommand\color[2][]{}%
  }%
  \providecommand\transparent[1]{%
    \errmessage{(Inkscape) Transparency is used (non-zero) for the text in Inkscape, but the package 'transparent.sty' is not loaded}%
    \renewcommand\transparent[1]{}%
  }%
  \providecommand\rotatebox[2]{#2}%
  \newcommand*\fsize{\dimexpr\f@size pt\relax}%
  \newcommand*\lineheight[1]{\fontsize{\fsize}{#1\fsize}\selectfont}%
  \ifx\svgwidth\undefined%
    \setlength{\unitlength}{677.05385365bp}%
    \ifx\svgscale\undefined%
      \relax%
    \else%
      \setlength{\unitlength}{\unitlength * \real{\svgscale}}%
    \fi%
  \else%
    \setlength{\unitlength}{\svgwidth}%
  \fi%
  \global\let\svgwidth\undefined%
  \global\let\svgscale\undefined%
  \makeatother%
  \begin{picture}(1,1.17545797)%
    \lineheight{1}%
    \setlength\tabcolsep{0pt}%
    \put(0,0){\includegraphics[width=\unitlength,page=1]{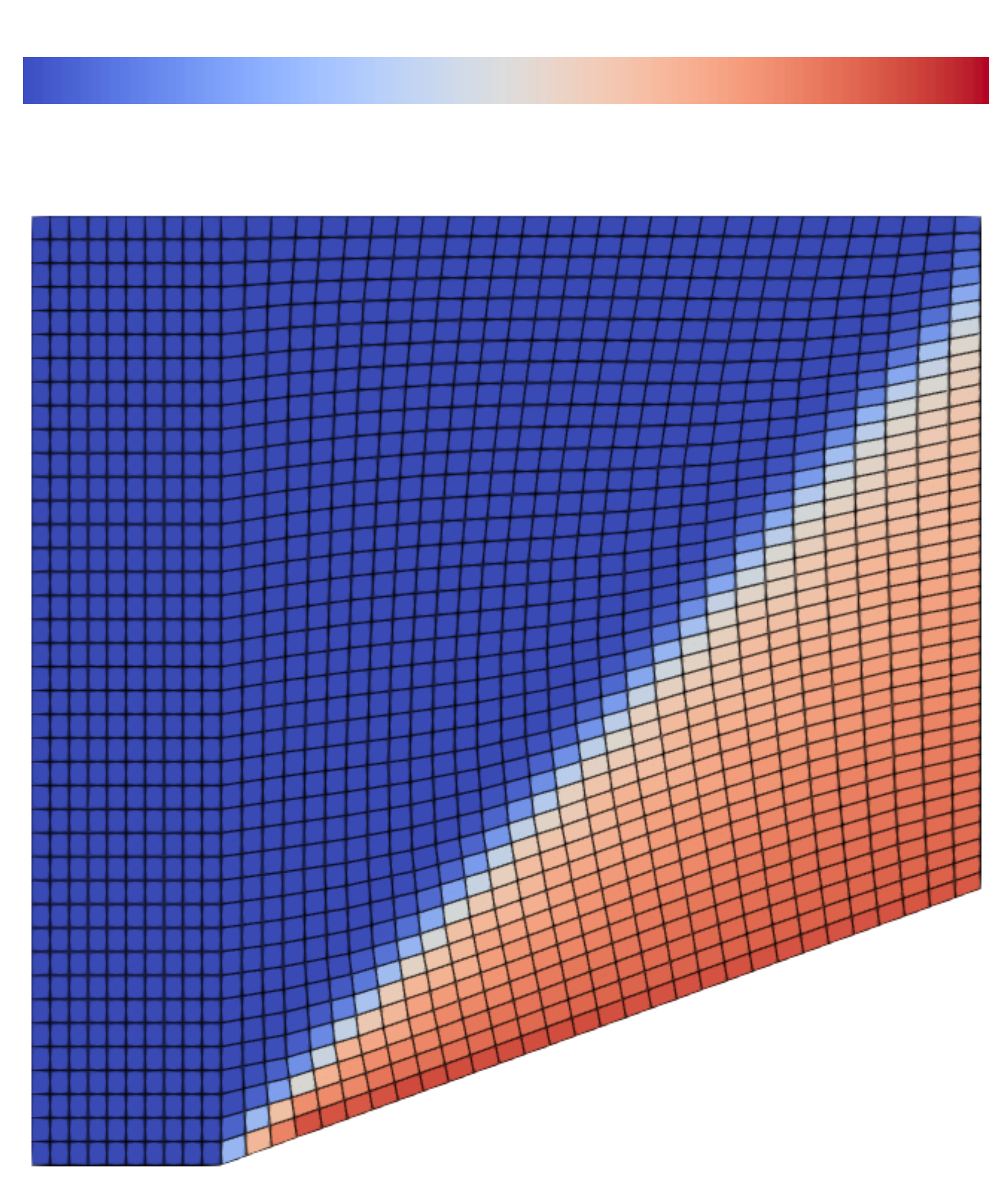}}%
    \put(0.98133679,1.00197991){\color[rgb]{0,0,0}\makebox(0,0)[t]{\lineheight{1.25}\smash{\begin{tabular}[t]{c}1.1\end{tabular}}}}%
    \put(0.7405878,1.00197991){\color[rgb]{0,0,0}\makebox(0,0)[t]{\lineheight{1.25}\smash{\begin{tabular}[t]{c}1.2\end{tabular}}}}%
    \put(0.50131584,1.00197991){\color[rgb]{0,0,0}\makebox(0,0)[t]{\lineheight{1.25}\smash{\begin{tabular}[t]{c}1.3\end{tabular}}}}%
    \put(0.26204431,1.00197991){\color[rgb]{0,0,0}\makebox(0,0)[t]{\lineheight{1.25}\smash{\begin{tabular}[t]{c}1.4\end{tabular}}}}%
    \put(0.02277218,1.00197991){\color[rgb]{0,0,0}\makebox(0,0)[t]{\lineheight{1.25}\smash{\begin{tabular}[t]{c}1.5\end{tabular}}}}%
    \put(0.50279283,1.14746315){\color[rgb]{0,0,0}\makebox(0,0)[t]{\lineheight{1.25}\smash{\begin{tabular}[t]{c}Mach Number\end{tabular}}}}%
    \put(0,0){\includegraphics[width=\unitlength,page=2]{cone20_M.pdf}}%
  \end{picture}%
\endgroup%

    \caption{Sharp-nosed 20$^\circ$ cone results: Mach number}
    \label{cone20_M}
\end{figure}

This multi-stage workflow is partially a legacy of the code's early history, when stand-alone grid generation tools did not exist and users had to specify their own flow domain in precise detail, but the process continues to have many benefits. The built-in grid generation is useful for making parametric grids, and the programmable input allows the user to compute interesting and complicated things in their setup process, such as a spatially varying initial flowfield or boundary conditions that depend on an in-situ compressible flow calculation. The verbose configuration files are also useful for documentation and reproducibility purposes, essentially serving as an automatically produced log of the simulation settings used in any research work.

\FloatBarrier
\subsection{Free Piston Driver}
The earliest calculations done with Eilmer's predecessors involved simulating expansion tunnels, large machines which are used to generate controlled hypersonic flow for laboratory experiments. These facilities typically use a heavy piston to compress a large slug of driver gas, which eventually bursts through a thick steel diaphragm and drives an incident shockwave through the facility, heating and pressurising the test gas that will eventually flow over the model. 

\begin{figure}[h]
    \tiny
    \def\svgwidth{0.9\columnwidth}
    \input{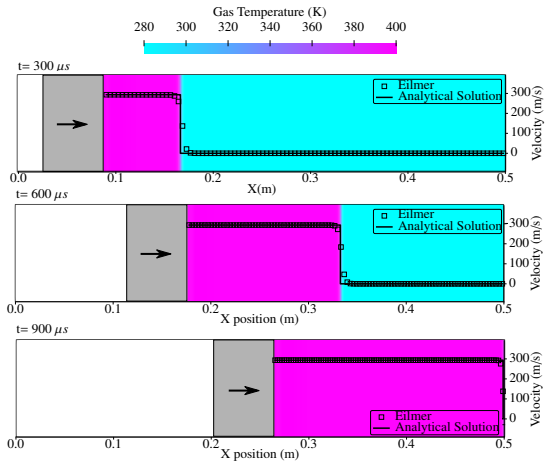}
    \caption{Free piston driver results: Temperature (K) and X-velocity (m/s)}
    \label{piston_T}
\end{figure}

Numerical simulations are used extensively to configure the tunnels before an experiment and to interpret the results afterward, and the modern Eilmer code has many capabilities of interest to experimenters, including a moving mesh transient solver which is the subject of this example. Figure \ref{piston_T} shows the results of a simple calculation where a piston is driven at constant velocity through an axisymmetric tube. The motion drives a shockwave through the gas with an analytically calculable velocity that compares well to the numerical results. 

\FloatBarrier
\subsection{Re-Entry Capsule}
High fidelity CFD is also used extensively in the design of the re-entry capsules that return people and equipment from space. These capsules are equipped with a sacrificial ablative heat-shield that absorbs some of the ferocious heating applied to the spacecraft's windward flank, burning away in layers and ejecting the absorbed energy away from the vehicle. Designing these heat shields requires a reasonably precise estimate of the peak and integrated heat transfer rates, since a too-thin layer of ablative material will not protect the vehicle properly, but a too-thick one will take up precious mass that could have been used for the payload. For these reasons, a large amount of research has gone into developing sophisticated engineering models for re-entry flows and providing experimental datasets they can be validated against. The application in this section is one such validation exercise, a subscale Apollo capsule model at conditions matching the second Project FIRE flight test from May 1965, specifically the 1636 second trajectory point at 71 kilometres altitude, where the model was travelling at 11.2 kilometres per second. The flow is modelled using a viscous, two-temperature, 11-species air model based on \citet{gnoffo_1989}, with reactions and thermal relaxation rates from \citet{park_review_I}. A 2D axisymmetric structured grid with 100x257 elements is used, with geometric clustering applied to the wall to achieve a first cell height of 2 $\mu m$, enough to grid-converge the wall heat transfer. 

\begin{figure}[h]
    \centering
    \scriptsize
    \def\svgwidth{0.65\columnwidth}
\begingroup%
  \makeatletter%
  \providecommand\color[2][]{%
    \errmessage{(Inkscape) Color is used for the text in Inkscape, but the package 'color.sty' is not loaded}%
    \renewcommand\color[2][]{}%
  }%
  \providecommand\transparent[1]{%
    \errmessage{(Inkscape) Transparency is used (non-zero) for the text in Inkscape, but the package 'transparent.sty' is not loaded}%
    \renewcommand\transparent[1]{}%
  }%
  \providecommand\rotatebox[2]{#2}%
  \newcommand*\fsize{\dimexpr\f@size pt\relax}%
  \newcommand*\lineheight[1]{\fontsize{\fsize}{#1\fsize}\selectfont}%
  \ifx\svgwidth\undefined%
    \setlength{\unitlength}{803.3179476bp}%
    \ifx\svgscale\undefined%
      \relax%
    \else%
      \setlength{\unitlength}{\unitlength * \real{\svgscale}}%
    \fi%
  \else%
    \setlength{\unitlength}{\svgwidth}%
  \fi%
  \global\let\svgwidth\undefined%
  \global\let\svgscale\undefined%
  \makeatother%
  \begin{picture}(1,1.26539844)%
    \lineheight{1}%
    \setlength\tabcolsep{0pt}%
    \put(0,0){\includegraphics[width=\unitlength,page=1]{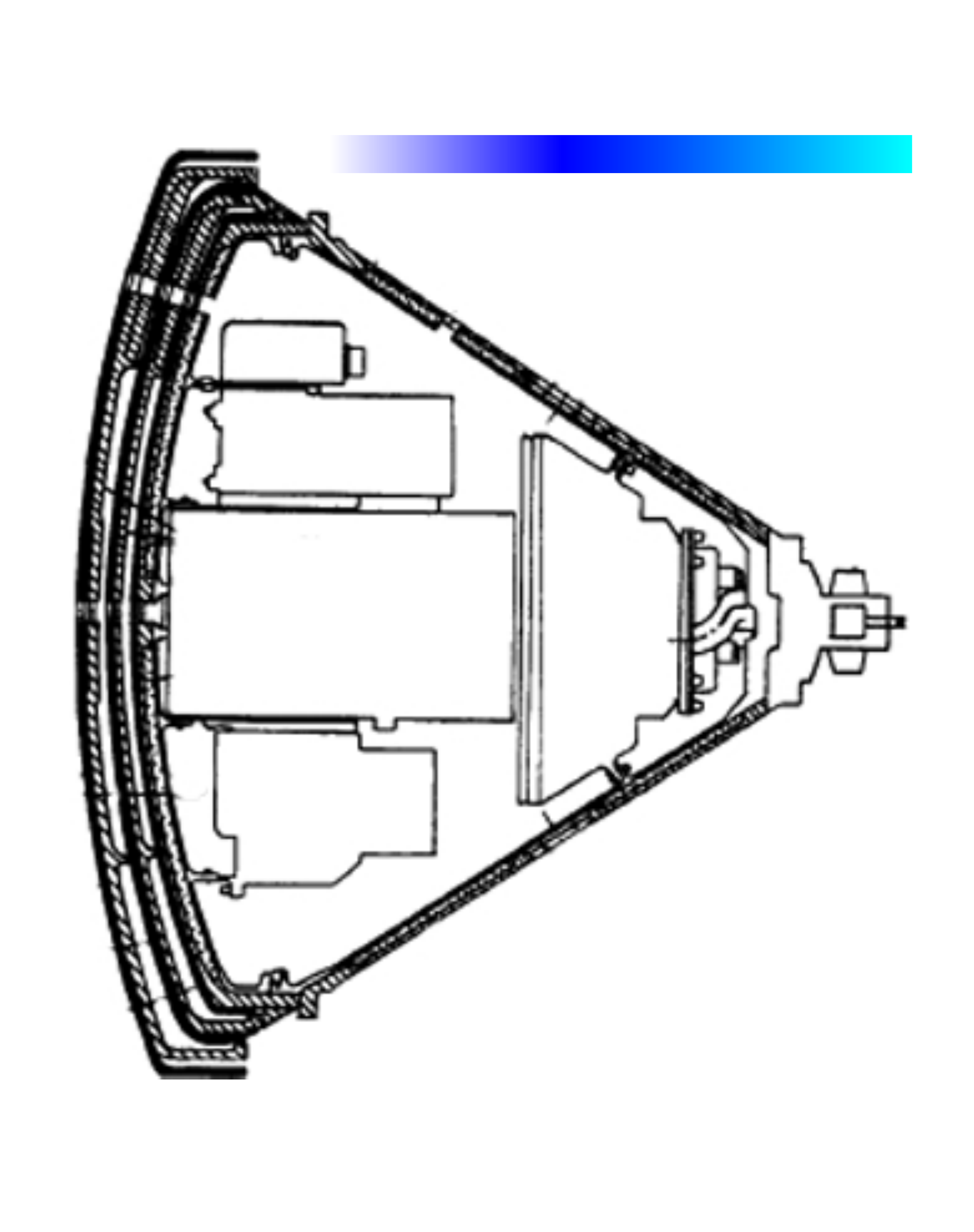}}%
    \put(0.33912383,1.16757968){\color[rgb]{0,0,0}\makebox(0,0)[t]{\lineheight{1.25}\smash{\begin{tabular}[t]{c}0\end{tabular}}}}%
    \put(0.53829782,1.16757968){\color[rgb]{0,0,0}\makebox(0,0)[t]{\lineheight{1.25}\smash{\begin{tabular}[t]{c}4000\end{tabular}}}}%
    \put(0.73747178,1.16757968){\color[rgb]{0,0,0}\makebox(0,0)[t]{\lineheight{1.25}\smash{\begin{tabular}[t]{c}8000\end{tabular}}}}%
    \put(0.93664576,1.16757968){\color[rgb]{0,0,0}\makebox(0,0)[t]{\lineheight{1.25}\smash{\begin{tabular}[t]{c}12,000\end{tabular}}}}%
    \put(0,0){\includegraphics[width=\unitlength,page=2]{fireII_tve.pdf}}%
    \put(0.30873973,1.22391619){\color[rgb]{0,0,0}\makebox(0,0)[lt]{\lineheight{1.25}\smash{\begin{tabular}[t]{l}Vibrational/Electronic Temperature (K)\end{tabular}}}}%
    \put(0,0){\includegraphics[width=\unitlength,page=3]{fireII_tve.pdf}}%
  \end{picture}%
\endgroup%

    \caption{Vibrational/Electronic temperature flowfield over a scaled Apollo capsule. Capsule geometry from Ref. \cite{liechty_dsmc11}, Figure 1.}
    \label{fireII_tve}
\end{figure}

\begin{figure}[h]
    \scriptsize
    \def\svgwidth{1.0\columnwidth}
    \input{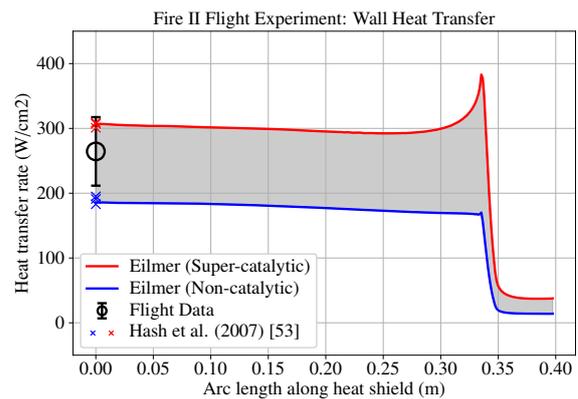}
    \caption{Wall heat transfer data from simulation of Project FIRE, flight II.}
    \label{fireII_wht}
\end{figure}

The biggest modelling uncertainty in the problem comes from the impact of surface chemistry. Figure \ref{fireII_wht} shows the predicted heat transfer rates for two nearly identical simulations, one with a non-catalytic wall where the wall-normal species density gradient is set to zero, and a second with a super-catalytic wall where the actual mass fractions are specified to be the same as the freestream. (This is effectively the same as a fully catalytic wall, given the surface temperature at these conditions is set to 810 K). At the stagnation point, Eilmer predicts a wall heat transfer rate of 185.6 $W/cm^2$ for the non-catalytic simulation and 306.6 $W/cm^2$ for the catalytic one, which neatly brackets the actual convective heating rate of 264.5$W/cm^2$, computed by subtracting the measured radiative heating from the total rate.

The computed values are also in good agreement with the simulation results presented in \citet{hash_fireii07}, a code comparison between NASA Ames's DPLR code, NASA Langley's LAURA, and the University of Minnesota's US3D, all using very similar physical models to those implemented in Eilmer.

\FloatBarrier
\subsection{Aerodynamic Shape Optimisation}
A single CFD solution of the flow around an object can predict values of lift, drag and heat transfer that an engineer can use in designing a supersonic aircraft. But one of the most interesting frontier of numerical research involves coupling the solver with a numerical optimisation technique to produce \emph{automated} designs that maximise performance subject to constraints. This subsection is devoted to an example of this workflow using Eilmer, taken from Kyle Damm's PhD thesis \cite{damm_phd}. The exercise is a simple proof-of-concept optimisation of a slender axisymmetric body, which is known to have a curved shape that minimises the total wave drag. The body shape is controlled by a 20-point B\'{e}zier curve with fixed ends, and the open-source optimisation toolkit DAKOTA \cite{dakota} is used to shift the B\'{e}zier control points around until a minimum drag configuration is found.

\begin{figure}[h]
    \tiny
    \def\svgwidth{1.0\columnwidth}
\begingroup%
  \makeatletter%
  \providecommand\color[2][]{%
    \errmessage{(Inkscape) Color is used for the text in Inkscape, but the package 'color.sty' is not loaded}%
    \renewcommand\color[2][]{}%
  }%
  \providecommand\transparent[1]{%
    \errmessage{(Inkscape) Transparency is used (non-zero) for the text in Inkscape, but the package 'transparent.sty' is not loaded}%
    \renewcommand\transparent[1]{}%
  }%
  \providecommand\rotatebox[2]{#2}%
  \newcommand*\fsize{\dimexpr\f@size pt\relax}%
  \newcommand*\lineheight[1]{\fontsize{\fsize}{#1\fsize}\selectfont}%
  \ifx\svgwidth\undefined%
    \setlength{\unitlength}{1422.11218109bp}%
    \ifx\svgscale\undefined%
      \relax%
    \else%
      \setlength{\unitlength}{\unitlength * \real{\svgscale}}%
    \fi%
  \else%
    \setlength{\unitlength}{\svgwidth}%
  \fi%
  \global\let\svgwidth\undefined%
  \global\let\svgscale\undefined%
  \makeatother%
  \begin{picture}(1,0.38783389)%
    \lineheight{1}%
    \setlength\tabcolsep{0pt}%
    \put(0,0){\includegraphics[width=\unitlength,page=1]{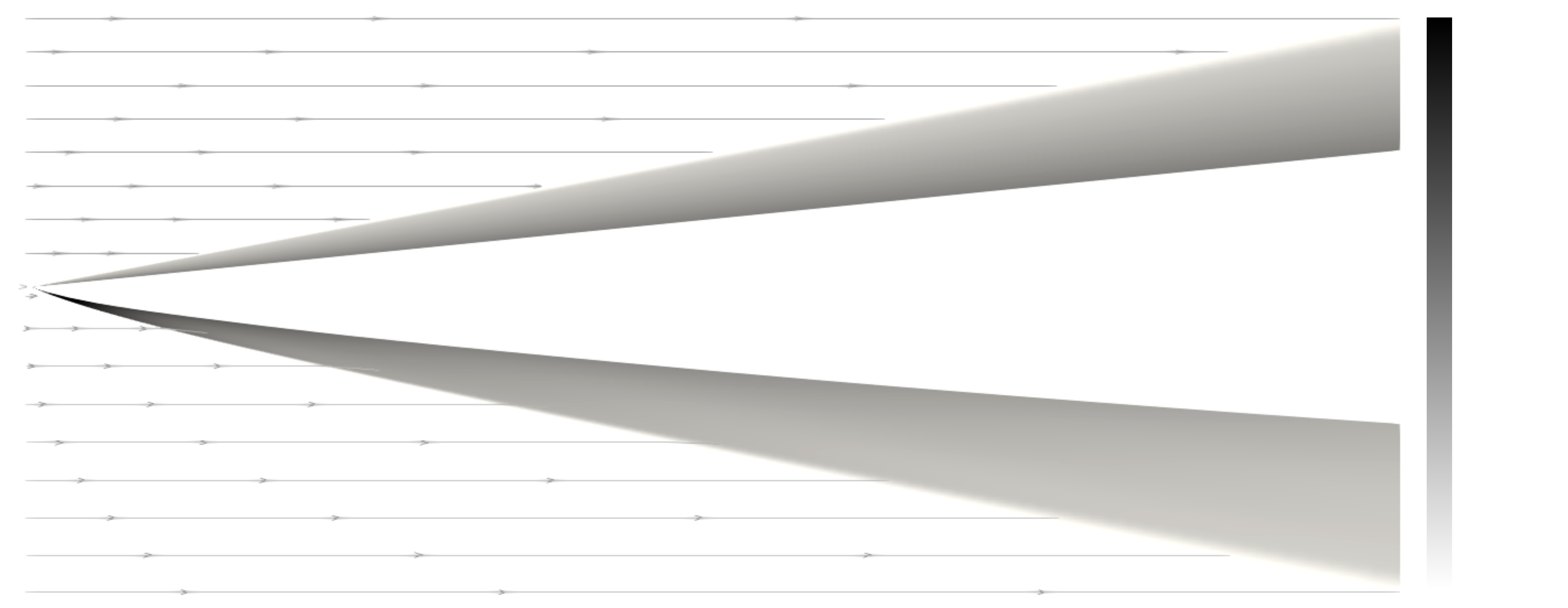}}%
    \put(0.94023522,0.01514255){\color[rgb]{0,0,0}\makebox(0,0)[lt]{\lineheight{1.25}\smash{\begin{tabular}[t]{l}0\end{tabular}}}}%
    \put(0.94023522,0.10585269){\color[rgb]{0,0,0}\makebox(0,0)[lt]{\lineheight{1.25}\smash{\begin{tabular}[t]{l}100\end{tabular}}}}%
    \put(0.94023522,0.19726602){\color[rgb]{0,0,0}\makebox(0,0)[lt]{\lineheight{1.25}\smash{\begin{tabular}[t]{l}200\end{tabular}}}}%
    \put(0.94023522,0.28797616){\color[rgb]{0,0,0}\makebox(0,0)[lt]{\lineheight{1.25}\smash{\begin{tabular}[t]{l}300\end{tabular}}}}%
    \put(0.94023522,0.37938949){\color[rgb]{0,0,0}\makebox(0,0)[lt]{\lineheight{1.25}\smash{\begin{tabular}[t]{l}400\end{tabular}}}}%
    \put(0.99632097,0.19726602){\color[rgb]{0,0,0}\rotatebox{90}{\makebox(0,0)[t]{\lineheight{1.25}\smash{\begin{tabular}[t]{c}Y-Velocity (m/s)\end{tabular}}}}}%
    \put(0,0){\includegraphics[width=\unitlength,page=2]{wedge_composite.pdf}}%
  \end{picture}%
\endgroup%

    \caption{Before (blue) and after (red) axisymmetric wedge subjected to optimisation for minimum wave drag.}
    \label{wedge_composite}
\end{figure}

Figure \ref{wedge_composite} shows the before and after results of the optimisation, with the initial shape (a right circular cone) shown in blue, the optimised curve shown in red, and the steady flowfield of each state shown as the black-on-white colourmap. The optimiser has induced a subtle curvature to the surface and reduced the wave drag by approximately $7 \%$, using 66 iterations of the optimisation loop. Crucially, the gradient of the objective function with respect to the B\'{e}zier control points is evaluated using an adjoint method that is built into the flow solver. Adjoint methods are crucial to CFD-based optimisation, as they allow the objective function gradients to be computed with just one CFD simulation, regardless of how many design variables are present. A more complete explanation of this methodology can be found in \citet{damm_adjoint20}, which applies the same method to a hypersonic vehicle inlet and presents more detail about the adjoint solver and the simulations involved. 

\FloatBarrier
\subsection{Double Cone Flow Physics Investigation}
\begin{figure}[h]
    \tiny
    \def\svgwidth{1.0\columnwidth}
    \input{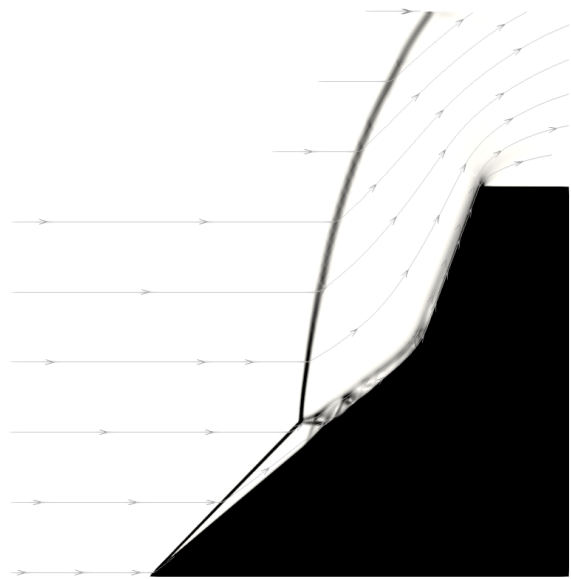}
    \caption{Steady flow from Ref. \cite{hornung_doublecones21}, $\theta_1=40^\circ$, $\theta_2=70^\circ$, condition A.}
    \label{rhograd_4070}
\end{figure}

Another popular application for flow simulation codes is doing fundamental research in theoretical fluid mechanics. Numerical simulations have a number of advantages over brick-and-mortar wind tunnels, and at least in laminar flow they can usually be relied on to generate exact and often quite complex solutions of the Navier-Stokes equations. An example is the recent paper of \citet{hornung_doublecones21}, where Eilmer was used to solve the high Mach-number flow over a large number of double-cones with different angles, producing either a stable or unstable shock structure in each case depending on the geometry. An example of the stable structure is shown in figure \ref{rhograd_4070}.

This figure is a numerical shadowgraph showing the density gradients over a double cone with angles of $\theta_1=40^\circ$ and $\theta_2=70^\circ$, with a flow Mach number of 7.7 and a composition of pure $N_2$. The shock structure is macroscopically steady, consisting of a conical oblique shock attached to the tip of the first cone and a complex triple point where it intersects with a supersonic jet and a small separation close to the wall.  

\begin{figure}[h]
    \tiny
    \def\svgwidth{1.0\columnwidth}
    \input{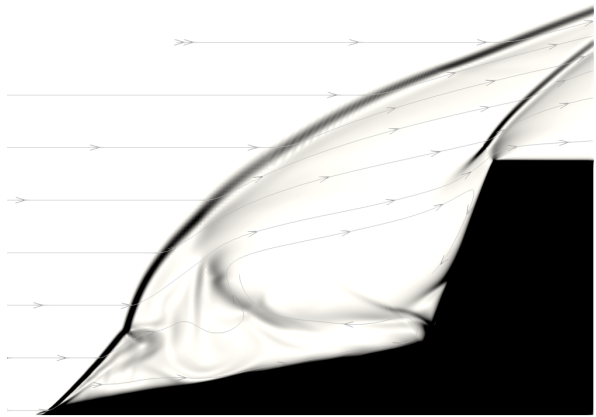}
    \caption{Unsteady flow from Ref. \cite{hornung_doublecones21}, $\theta_1=10^\circ$, $\theta_2=70^\circ$, condition A.}
    \label{rhograd_1070}
\end{figure}

Decreasing the angle of the first ramp strengthens the second shock in comparison to the first one, driving the separation point forward and potentially causing it to break free from the wall. At $\theta_1=0^\circ$ the geometry would essentially be a blunt body with a detached bow shock, but at intermediate values of $\theta_1$ the shock is neither attached nor detached but instead oscillates violently back and forth. A single frame of this process is shown in figure \ref{rhograd_1070}, taken from a simulation with $\theta_1=10^\circ$ and $\theta_2=70^\circ$, but otherwise using the same flow conditions as figure \ref{rhograd_4070}. The shock structure is in the act of surging forward, producing a large region of chaotic separated flow, and will shortly detach from the tip of the first cone and then reverse direction, eventually crashing back against the wall and beginning a new cycle of unsteadiness. The simulation captures multiple cycles of this unsteadiness using Eilmer's time accurate transient solver, and \citet{hornung_doublecones21} use many simulations over a wide parameter sweep to infer a boundary in $\theta_1$ vs. $\theta_2$ space where the instability begins.

\FloatBarrier
\subsection{Eilmer in the Classroom}
Throughout its history Eilmer has been used for both research and teaching at a handful of different universities. Many postgraduate students use the code in their research projects, and the codebase is full of small contributions from students who have have needed some cutting-edge simulation capability and managed to implement it themselves. This is one of the major advantages of using open-source software for academic research, and sparing students the fearsome learning curve of C++ was one of the key drivers for our adoption of the D programming language for the code's 2016-era rewrite.

At The University of Queensland, Eilmer is also used in our undergraduate and masters level course in computational fluid dynamics. This last example is a simple validation exercise performed by our final year mechanical and aerospace engineering students each year, in which they must match the experimentally measured shock standoff around a blunt cone immersed in a Mach 4 test flow generated by a small shock tunnel. The students do the experiments, collect high-speed Schlieren video of the models, build grids, run the simulations, and are responsible for extracting the shape of the shockwaves and accounting for the inevitable complications that arise in any real experiment. 
\begin{figure}[h]
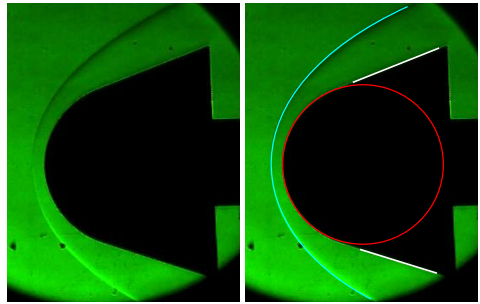

    \tiny
    \centering
    \def\svgwidth{0.39\columnwidth}
    \input{frame_19_clean.pdf_tex}
    \def\svgwidth{0.39\columnwidth}
    \input{frame_19.pdf_tex}
    \caption{Steady flow Schlieren image from 2021 MECH4480 student experiments}
    \label{frame_19}
\end{figure}

Figure \ref{frame_19} is an example of the experimental data, a single Schlieren image of a $15 mm$ radius cone with a half angle of $20^\circ$, extracted during the approximately one millisecond window of steady test time. A circle with a radius of 53 pixels has been fitted to the body, which can be used to determine the scale of the image and the position of the model. A cubic B\'{e}zier curve has been fitted to the shock shape, which suggests an experimental shock standoff distance of $\approx 2.2 mm$. 

\begin{figure}[h]
    \tiny
    \centering
    \def\svgwidth{0.70\columnwidth}
    \input{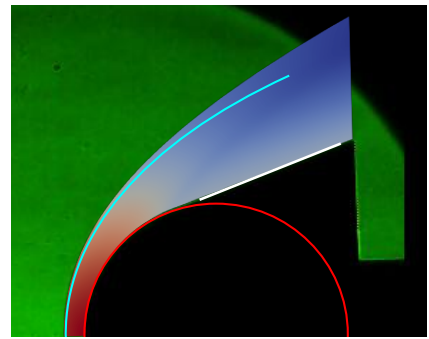}
    \caption{Shock fitted simulation temperature field overlayed on Schlieren results}
    \label{frame_19_overlay}
\end{figure}

Figure \ref{frame_19_overlay} is the same data but with a shock-fitted thermally perfect simulation of the experiment. 

The simulation predicts a shock standoff of $\approx 2.4 mm$ and is a reasonably good match to the data close to the stagnation point, though the simulated shape wanders above the measurements further downstream. This is most likely due to the unmodelled divergence in the nozzle's core-flow, which is assumed to be uniform in the simulation to avoid introducing undue complexity into the problem. In fact, handling this kind of mismatch between experiment and simulation is precisely what makes the exercise a great learning experience for the students.

\section{Conclusions}
\label{sec:conclusion}
This paper has been written to fulfil multiple purposes. For the broad aerospace research community, it gives a brief overview of the open-source compressible flow simulation codes available, both niche and comprehensive, and explains how Eilmer fits into the landscape. For users and future developers of the code, it lays out the basic mathematical principles and governing equations that are solved to produce the simulation data they rely on. A good understanding of these principles is critical for developers, and even casual users should have some appreciation of the code's fundamentals to avoid wandering into areas where their results may be invalid. For the developers of other scientific codes, we have tried to document the programming principles, testing tools, and Verification and Validation exercises that go on behind-the-scenes of the project. We have found that a small investment in diligent software engineering in the present can avoid large amounts of extra work in the future, and wish to join other advocates of adopting programming best-practice ideas in scientific computing. For users who work with HPC facilities we have included a section on parallelisation, with an example of a scaling exercise that should serve as a blueprint for others to follow. This should be particularly helpful to those applying for scarce supercomputing time, whose gatekeepers are increasingly anxious to know that their allocations are being used efficiently. Finally, for the general user who wants to know what Eilmer is capable of, we have included a handful of interesting example simulations that cover theoretical and applied fluid mechanics research, hypersonic engineering calculations, and teaching exercises. 

All of these purposes are really subgoals of the main goal of this paper, which is to share our project with the world. One of the great triumphs of the modern internet is the open-source software movement, which has strived to make computer programs that are \emph{free}: Both in the sense of \emph{free beer} (they do not cost any money), and free as in \emph{freedom} (the source code is available for anyone to download, inspect, and modify). This philosophy overlaps with the broader mission of scientific research, which aims to discover information about the world that can be distributed to the benefit of everyone, and the proliferation of open-source research codes is the result of a harmonious match between the two ideas. By publishing both our code and this paper, we hope to join in this quiet revolution, and by helping others advance the march of scientific progress, to play a small part in a brighter future.





\bibliography{references}

\newpage
\newpage

\begin{figure*}[t]
{\large
\center
\textbf{Appendix 1: Manufactured Solution for 3D Viscous, Turbulent Flow}
}
\begin{equation*}
\rho = 1 - \frac{\sin{\left(\frac{9 \pi y}{20 L} \right)}}{10} + \frac{\sin{\left(\frac{4 \pi z}{5 L} \right)}}{10} + \frac{3 \sin{\left(\frac{\pi x z}{2 L^{2}} \right)}}{25} + \frac{3 \cos{\left(\frac{3 \pi x}{4 L} \right)}}{20} + \frac{2 \cos{\left(\frac{13 \pi x y}{20 L^{2}} \right)}}{25} + \frac{\cos{\left(\frac{3 \pi y z}{4 L^{2}} \right)}}{20}
\end{equation*}
\begin{equation*}
u = 70 + 7 \sin{\left(\frac{\pi x}{2 L} \right)} - 4 \sin{\left(\frac{9 \pi x z}{10 L^{2}} \right)} - 15 \cos{\left(\frac{17 \pi y}{20 L} \right)} - 10 \cos{\left(\frac{2 \pi z}{5 L} \right)} + 7 \cos{\left(\frac{3 \pi x y}{5 L^{2}} \right)} + 4 \cos{\left(\frac{4 \pi y z}{5 L^{2}} \right)}
\end{equation*}
\begin{equation*}
v = 90 - 5 \sin{\left(\frac{4 \pi x}{5 L} \right)} + 5 \sin{\left(\frac{3 \pi x z}{5 L^{2}} \right)} + 10 \cos{\left(\frac{4 \pi y}{5 L} \right)} + 5 \cos{\left(\frac{\pi z}{2 L} \right)} - 11 \cos{\left(\frac{9 \pi x y}{10 L^{2}} \right)} - 5 \cos{\left(\frac{2 \pi y z}{5 L^{2}} \right)}
\end{equation*}
\begin{equation*}
w = 80 +  10 \sin{\left(\frac{9 \pi y}{10 L} \right)} - 12 \sin{\left(\frac{2 \pi x y}{5 L^{2}} \right)} + 5 \sin{\left(\frac{3 \pi x z}{4 L^{2}} \right)} - 10 \cos{\left(\frac{17 \pi x}{20 L} \right)} + 12 \cos{\left(\frac{\pi z}{2 L} \right)} + 11 \cos{\left(\frac{4 \pi y z}{5 L^{2}} \right)}
\end{equation*}
\begin{equation*}
p = 100,000 +  20,000 \sin{\left(\frac{17 \pi z}{20 L} \right)} + 10,000 \sin{\left(\frac{4 \pi x z}{5 L^{2}} \right)} + 20,000 \cos{\left(\frac{2 \pi x}{5 L} \right)} + 50,000 \cos{\left(\frac{9 \pi y}{20 L} \right)} - 25,000 \cos{\left(\frac{3 \pi x y}{4 L^{2}} \right)} - 10,000 \cos{\left(\frac{7 \pi y z}{10 L^{2}} \right)}
\end{equation*}
\begin{equation*}
\hat{\nu} = 1 + \frac{4 \sin{\left(\frac{4 \pi z}{5 L} \right)}}{5} - \frac{3 \sin{\left(\frac{3 \pi x z}{5 L^{2}} \right)}}{5} + \frac{6 \cos{\left(\frac{7 \pi x}{20 L} \right)}}{25} - \frac{3 \cos{\left(\frac{2 \pi y}{5 L} \right)}}{10} + \frac{3 \cos{\left(\frac{\pi x y}{2 L^{2}} \right)}}{4} + \frac{\cos{\left(\frac{\pi y z}{4 L^{2}} \right)}}{2}
\end{equation*}
\label{mms_appendix}
\end{figure*}







\end{document}